\documentclass[preprint, authoryear, 5p, number]{elsarticle}
\pdfoutput=1
\usepackage{units}
\usepackage{amsmath}
\usepackage{amssymb}
\usepackage[english]{babel}
\usepackage{color}
\usepackage{array}
\usepackage{booktabs}
\usepackage{graphicx}
\usepackage{array}
\usepackage{gensymb}
\usepackage{relsize}
\usepackage[latin1]{inputenc}
\usepackage{float}
\usepackage{hyperref}
\usepackage{lscape}
\usepackage{csquotes}
\usepackage{rotating}
\usepackage{tabularx}
\usepackage{multicol}
\usepackage{lipsum}
\usepackage{listings}
\usepackage{multicol}
\usepackage{wrapfig}
\usepackage{kantlipsum}
\usepackage{lscape}
\usepackage{rotating}
\usepackage{epstopdf}
\def\CC{{C\nolinebreak[4]\hspace{-.05em}\raisebox{.4ex}{\tiny\bf ++}}}

\DeclareFixedFont{\ttb}{T1}{txtt}{bx}{n}{12} 
\DeclareFixedFont{\ttm}{T1}{txtt}{m}{n}{12}  

\usepackage{color}
\definecolor{deepblue}{rgb}{0,0,0.5}
\definecolor{deepred}{rgb}{0.6,0,0}
\definecolor{deepgreen}{rgb}{0,0.5,0}

\newcommand\pythonstyle{\lstset{
language=Python,
basicstyle=\ttm,
otherkeywords={self},             
keywordstyle=\ttb\color{deepblue},
emph={MyClass,__init__},          
emphstyle=\ttb\color{deepred},    
stringstyle=\color{deepgreen},
frame=tb,                         
showstringspaces=false            %
}}

\lstnewenvironment{python}[1][]
{
\pythonstyle
\lstset{#1}
}
{}

\newcommand\pythoninline[1]{{\pythonstyle\lstinline!#1!}}

\journal{Astronomy and Computing}
 
\begin{document}
 
\begin{frontmatter}
 
\title{SlicerAstro: a 3-D interactive visual analytics tool for H\,{\large I} data}

 \author[a]{D.~Punzo\corref{cor1}}
 \ead{D.Punzo@astro.rug.nl}
 
 \author[a]{J.M.~van der Hulst}
 \author[b]{J.B.T.M.~Roerdink}
 \author[c]{J.C.~Fillion-Robin}
 \author[d,e]{L.~Yu}

 \cortext[cor1]{Corresponding author}

 \address[a]{Kapteyn Astronomical Institute, University of Groningen, Landleven 12, 9747 AD Groningen, The Netherlands}
  
 \address[b]{Johann Bernoulli Institute for Mathematics and Computer Science, University of Groningen, Nijenborgh 9, 9747 AG Groningen, The Netherlands}
 
 \address[c]{Kitware Inc, Clifton Park, NY, The United States}
 
 \address[d]{University Medical Center Groningen, Center for Medical Imaging North East Netherlands, University of Groningen, Hanzeplein 1, 9713 GZ Groningen, The Netherlands}
 
 \address[e]{Hangzhou Dianzi University, Zhejiang, China}

\begin{abstract}
SKA precursors are capable of detecting hundreds of galaxies in H\,{\small I} in a 
single 12 hours pointing. In deeper surveys one will probe more easily faint 
H\,{\small I}
structures, typically located in the vicinity of galaxies, such as tails, filaments,
and extraplanar gas. The importance of interactive visualization in data exploration has been demonstrated by the wide use of tools (e.g.~$\tt{Karma}$, $\tt{Casaviewer}$, $\tt{VISIONS}$) that help users to receive immediate feedback when manipulating the data.
We have developed $\tt{SlicerAstro}$, a 3-D
interactive viewer with new analysis capabilities, based on 
traditional 2-D input/output hardware. These capabilities enhance
the data inspection, allowing faster analysis of complex sources
than with traditional tools.
$\tt{SlicerAstro}$ is an open-source extension of $\tt{3DSlicer}$,  
a multi-platform open source software package for visualization and medical image processing.

We demonstrate the capabilities of the current stable binary release of $\tt{SlicerAstro}$, 
which offers the following features:
i) handling of FITS files and astronomical coordinate systems; ii) 
coupled 2-D/3-D visualization; iii) interactive filtering; 
iv) interactive 3-D masking; v) and interactive 3-D modeling.
In addition, $\tt{SlicerAstro}$ has been designed with
a strong, stable and modular \CC\; core, and its classes are also accessible via
$\tt{Python}$ scripting,
allowing great flexibility for user-customized visualization and analysis tasks.

\end{abstract}
 
 
\end{frontmatter}

\section{Introduction}\label{intro}

Upcoming neutral hydrogen (H\,{\small I}) surveys \citep[e.g.,][]{ Apertif3,duffy} will deliver large 
datasets. The daily data-flow will be of the order of TBytes
and several hundreds of galaxies will be detected.
To find and 
characterize H\,{\small I} objects, automated processing methods must use all of the three-dimensional 
(3-D) information (two positional dimensions and one spectral dimension) that the surveys make 
available. 

In this context, 3-D visualization techniques provide a powerful tool
to inspect the sources under study. 
In fact, the 3-D view of a galaxy simultaneously 
presents both its H\,{\small I} distribution and its 
kinematics providing an immediate overview of the structures and coherence 
in the data \citep{Oosterloo, Goodman, Punzo2015}.
In addition, user interaction
in the 3-D environment provides capabilities which astronomers can 
use to quickly analyze complex sources found by automated pipelines 
\citep[e.g., $\tt{Duchamp}$ and $\tt{SoFiA}$;][]{Whiting, sofia}. 
These sources include interacting galaxies, tidal tails, filaments, 
and stripped galaxies, and the majority
will not exceed dimensions greater than $10^8$ voxels\footnote{ Voxels are 3-D pixels.}. 

Performing interactive 3-D rendering (and analysis) of H\,{\small I} sources
is computationally affordable using a modern desktop \citep{Punzo2015}.
This has stimulated further development 
of 3-D visualization tools for astronomical purposes. For example, 
different package developments have recently been undertaken, exploiting: 
the rendering engine of $\tt{Blender}$
\footnote{\url{https://www.blender.org/}},
an open source software for 3-D animations 
$\,$\citep{frelled,KentBook,AstroBlend};
indirect volume rendering\footnote{In scientific visualization and computer graphics,
volume rendering is a set of techniques used
to display a 2-D projection of a 3-D discretely sampled dataset.}
available in the Visualization ToolKit, $\tt{VTK}$
\footnote{\url{http://www.vtk.org/}},
and $\tt{Mayavi2}$\footnote{\url{http://code.enthought.com/projects/mayavi/}} \citep{X3D}; 
stereoscopic visualization and 3-D interaction hardware using the gaming engine $\tt{Unity}$
\footnote{\url{https://unity3d.com/}}
\citep{Ferrand}; and a large-scale, hybrid
visualization and supercomputing environment \citep{Vohl}.
Although the previous packages have introduced 3-D rendering solutions to visualize 
3-D astronomical datasets, they do not fully satisfy our visualization requirements 
(see Section~\ref{design}).

In this paper, we present
$\tt{SlicerAstro}$ \footnote{\url{https://github.com/Punzo/SlicerAstro}} \citep{SlicerAstro},
an extension of 
$\tt{3DSlicer}$\footnote{\url{https://www.slicer.org/}} 
\citep[a multi-platform open source software package 
for visualization and medical image processing;][]{Slicer}, that aims to provide an
interactive 3-D visual analytics tool based on traditional
2-D input/output hardware.  

In Section~\ref{design} we describe the design of $\tt{SlicerAstro}$. 
In Section~\ref{filtering} we show how interactive filtering and 3-D visualization 
can boost the inspection of faint
complex sources. In Section~\ref{masking} we describe the interactive 3-D masking capabilities 
available in $\tt{SlicerAstro}$. In Section~\ref{modeling} we show how 3-D visualization, coupled with
interactive modeling, provides additional capabilities 
helping the discovery and analysis of subtle structures in the 3-D domain.
In Section~\ref{conclusion} we discuss the efficiency of such visual analytics techniques for
helping astronomers in the analysis of complex sources.

\section{The $\tt{SlicerAstro}$ environment}\label{design}

An exhaustive review of open-source 3-D visualization packages in \cite{Punzo2015} led to the choice of 
$\tt{3DSlicer}$ as the preferred platform for the development of $\tt{Slicer}$-$\tt{Astro}$. 
The most important deciding factors included the following:

\begin{enumerate}[I)]
\item $\tt{3DSlicer}$ is an open-source platform with a Berkeley
Software Distribution (BSD) license, which 
allows for free utilization of the software;
\item the software has a flexible environment for code development and collaboration;
\item $\tt{3DSlicer}$ has adequate documentation for both developers and users;
\item the $\tt{3DSlicer}$ software has a large number of active developers;
\item the $\tt{3DSlicer}$ interface already has numerous quantitative features 
      e.g., data probing, setting fiducial markups\footnote{A fiducial
      markup or fiducial is an object placed in the field of view of an imaging system 
      which appears in the image produced, for use as a point of reference or a measure.}
      and listing their position, 2-D/3-D rulers and calculating statistics in a selected volume).
\end{enumerate}

Several of the medical visualization tools present in $\tt{3DSlicer}$ 
suit the needs of astronomical applications. 
For example, $\tt{3DSlicer}$ optimizes the display layout and the process of 
navigating through data for parallel two-dimen-sional visualizations (e.g., movies of channel maps).

In addition, $\tt{3DSlicer}$ has been adopted by 
Kitware\footnote{\url{https://www.kitware.com/}} as key open-source platform similarly to 
$\tt{VTK}$, $\tt{ITK}$\footnote{\url{https://itk.org/}} and 
$\tt{Paraview}$\footnote{\url{http://www.paraview.org/}}
which Kitware has been supporting for more than 15 years. This guarantees 
long-term support and future updates of $\tt{3DSlicer}$.

\subsection{Design}\label{design}

\cite{Punzo2015} analyzed and reviewed the requirements for the visualization 
of H\,{\small I} in and around galaxies. These include handling  
the loading and writing of Flexible Image Transport System (FITS)
files \citep{Pence}, the ability to display
astronomical World Coordinates System \citep[WCS;][]{Calabretta, Greisen}, 
interactive 3-D high-quality rendering capabilities 
\citep[i.e., \textit{graphics processing unit} (GPU)-accelerated ray casting rendering][]{Roth, Schroeder}
and interactive linking between 1-D/2-D/3-D views. 
Interactive visualization which allows the user to 
extract quantitative information 
directly from the visual presentation
is also of primary importance: probing the data with a 
cursor; displaying coordinate axes in the 2-D views; 
performing 3-D segmentation\footnote{Image segmentation is the process of partitioning an
image into disjoint regions that are uniform with respect to some property.} techniques;
linked 1-D/2-D/3-D region of interest (ROI) selection and the ability to  
calculate statistics (e.g., mean, $rms$, maximum, minimum, etc.)
in a specific area or volume.
Another requirement is to couple analysis techniques
such as interactive smoothing and 
tilted-ring model fitting to visualization.
Therefore, comparative visualization (multiple views, overlaid visualizations, etc.)
is fundamental for comparing the raw data with the smoothed version and/or the models.
The last requirement is interoperability\footnote{Interoperability is 
the ability of different information technology systems and 
software applications to communicate, exchange data, and use the 
information that has been exchanged.} with virtual observatory (VO) tools \citep{samp}.
Moreover, in order to facilitate collaborative work, the source code must be 
open, modular, well documented, and well maintained.

The current version of the $\tt{3DSlicer}$ software provides several of these
capabilities: CPU and GPU rendering based on the $\tt{VTK}$, interface optimized 
for 2-D visualization with a high-level of linking between the 2-D and 3-D views,
2-D and 3-D segmentations techniques, high-level of modularity in the source code, 
embedded python console in the user interface for fast interaction with the 
$\tt{3DSlicer}$ application programming interface (API)\footnote{The API is a set of subroutine
definitions, protocols, and tools for building application software.}, presence of detailed 
documentation for both users and developers.
In addition, we made a number of contributions to the $\tt{3DSlicer}$ source: we added 
more types of units in the $\tt{3DSlicer}$ standards and factorized 
the $\tt{DataProbe}$ module and widgets that control 
the 2-D views to allow their customization by 
$\tt{3DSlicer}$ extensions.  

In addition, to fulfill the requirements, the following capabilities 
have to be added: 
\begin{enumerate}[I)]
\item proper visualization of astronomical data-cubes using the FITS data format;
\item enabling interactive smoothing in all three dimensions;
\item interactive 3-D selection of H\,{\small I} sources;
\item interactive H\,{\small I} data modeling coupled to visualization;
\item generation of flux density profiles and histograms of the voxel intensities;
\item introduction of the SAMP protocol to enable interoperability 
      with $\tt{Topcat}$ \citep{Topcat}, and other VO tools and catalogs.
\end{enumerate}

These software capabilities are particular to astronomical applications
and, therefore, it is optimal to implement them in an extension of $\tt{3DSlicer}$,
i.e.~$\tt{SlicerAstro}$, rather than in its core.

In the next sections we will discuss the implementation and deployment 
of such capabilities and use the H\,{\small I} emission
in and around WEIN069 \citep{Mpati}, a galaxy in a region in the sky
where a filament of the Perseus-Pisces Supercluster (PPScl) crosses the
plane of the Milky Way, as an example.

\subsection{Implementation}

The $\tt{3DSlicer}$ plug-in mechanism enables the
rapid development of custom modules in different programming
languages and for different levels of integration: 
\begin{enumerate}[1)]
\item  The \textit{command-line interface modules} are 
standalone executables with a limited input/output 
argument complexity (simple argument types and no user interaction).
\item The \textit{loadable modules} are plugins implemented in the \CC\;
language that are integrated tightly 
in the $\tt{3DSlicer}$ core software. These modules have 
access to all other $\tt{3DSlicer}$ core modules and the internals of the application 
and they can define custom, interactive graphical user interfaces. 
\item  The \textit{scripted modules} are written in the $\tt{Python}$ language.
These modules can be developed and 
modified without rebuilding or restarting $\tt{3DSlicer}$ and they have 
similar access to the application 
internals as loadable modules.
\end{enumerate}

All objects (volumetric images, surface models, transforms, etc.) in 
$\tt{3DSlicer}$ are stored in a hierarchical structure of nodes 
encoded in the Medical Reality Modeling Language (MRML). Each MRML 
node has its own list of custom attributes that can be used
to specify additional characteristics for the 
data object. This method of storage enables the modules to have 
access to the MRML tree, allowing new 
extensions to leverage existing processing and visualization 
functions without directly interfering with other modules.

In addition, $\tt{3DSlicer}$ and its extensions are developed
using a $\tt{CMake}$-based\footnote{\url{https://cmake.org/}}
build system which greatly helps the development, 
packaging and testing of multi-platform software.

\begin{figure}[!ht]
\centering
\includegraphics[width=0.48\textwidth]{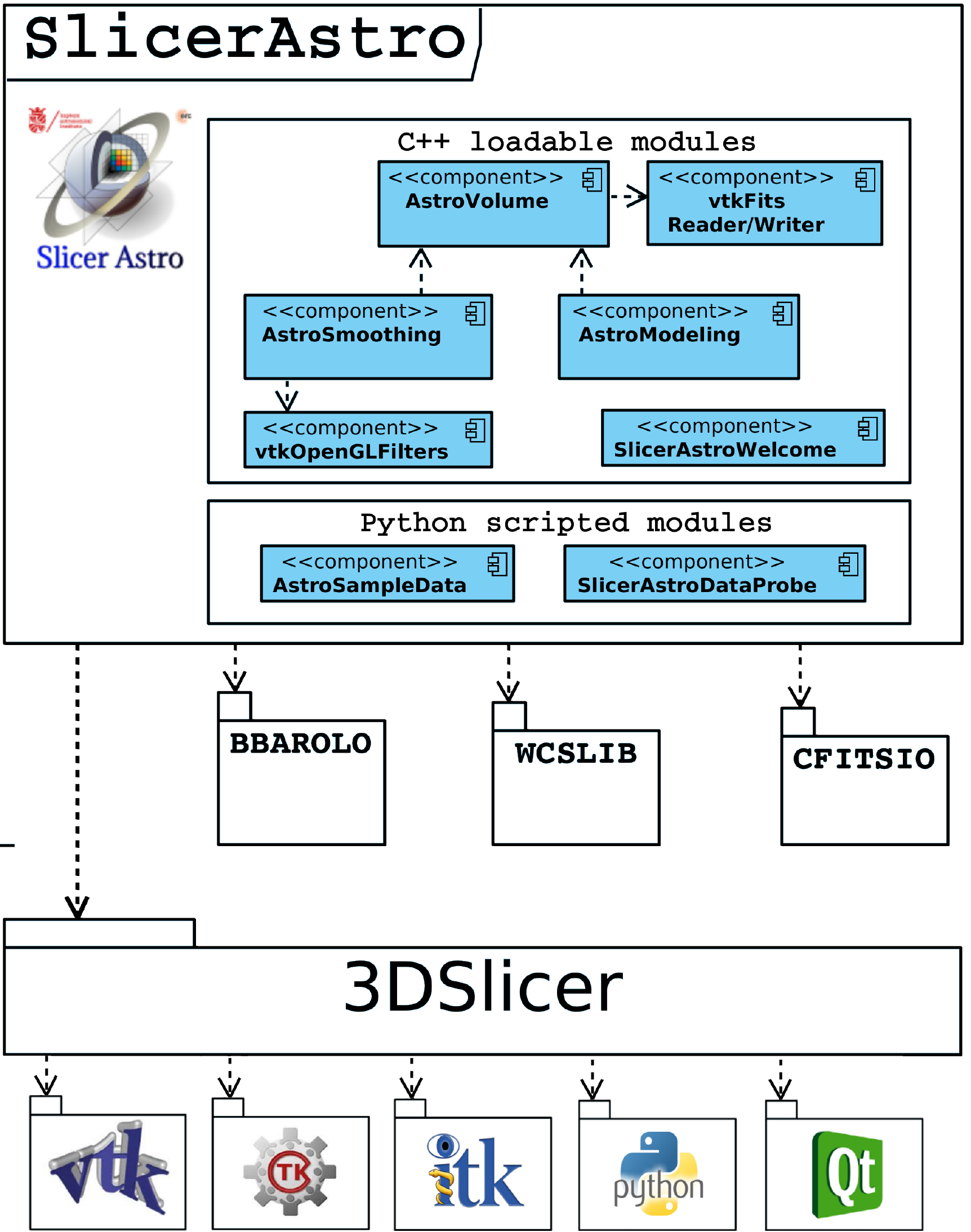}
\caption{The architecture of $\tt{SlicerAstro}$ is shown in the diagram.
The dashed arrows indicate the dependency of a component on another one.
The loadable modules are the main components of $\tt{SlicerAstro}$. 
The $\tt{AstroVolume}$ module is the core module of $\tt{SlicerAstro}$ 
and it provides an interface for
handling the loading and writing of FITS files, 
the control of the 2-D and 3-D color transfer functions,
and the display of the astronomical world coordinates system 
\citep[WCS;][]{Calabretta, Greisen}.
The $\tt{AstroSmoothing}$ and $\tt{AstroModeling}$ modules take care of
specific operations (smoothing and modeling respectively), with their own interface widgets
for user interaction.
The scripted modules have the role of utilities such as  
downloading sample datasets.
}
\label{SlicerAstroDesign}
\end{figure}  

The $\tt{SlicerAstro}$ functionality
is implemented as multiple plug-in modules, bundled as one downloadable 
extension. This modularization makes development and maintenance faster and affordable.
Moreover, extensions are built everyday for the nightly build of 
$\tt{3DSlicer}$ to identify breakage with the core.
The architecture of $\tt{SlicerAstro}$ is shown in Fig.~\ref{SlicerAstroDesign}. 
$\tt{SlicerAstro}$ uses 
the $\tt{CTK}$\footnote{\url{http://www.commontk.org}} and 
$\tt{Qt}$\footnote{\url{https://www.qt.io/}}
packages for user interface widgets, and the $\tt{VTK}$ 
library for the visualization (i.e.~2-D and 3-D 
rendering).
$\tt{SlicerAstro}$ depends also on: $\tt{CFITSIO}$ \citep{cfitsio}, $\tt{WCSLIB}$ 
\citep{wcslib} and $\tt{^{\rm3D}\,Barolo}$ \citep{bbarolo}.
The loadable modules are the main components of $\tt{SlicerAstro}$, while 
the scripted modules have the role of utilities such as presenting a welcoming interface and 
capabilities to download sample datasets.
The $\tt{AstroVolume}$ component is the core module (see Section~\ref{framework});
$\tt{AstroSmo}$-$\tt{othing}$ and $\tt{AstroModeling}$ modules take care of specific 
operations (smoothing and modeling respectively), with their own interface widgets
for user interaction (see Sections \ref{filtering} and \ref{modeling}).

$\tt{SlicerAstro}$ development focuses on H\,{\small I} datasets. 
Therefore, we currently provide modules which are mainly aimed for the 
analysis of H\,{\small I} data-cubes.
However, $\tt{SlicerAstro}$ can potentially enhance also the inspection of other 
datasets such as mm/submm molecular line data and optical integral field spectroscopic 
data. We will elaborate more in the potential of $\tt{SlicerAstro}$ for such datasets 
in Section~\ref{conclusion}.

\subsection{Interface framework}\label{framework}
The $\tt{AstroVolume}$ module provides an interface for
handling the loading and writing of FITS files, MRML nodes that 
store the data in the $\tt{3DSlicer}$ object-tree,
the display of the WCS and the control of the 2-D and 3-D color transfer functions.
 
\begin{figure}[!ht]
\centering
\includegraphics[width=0.48\textwidth]{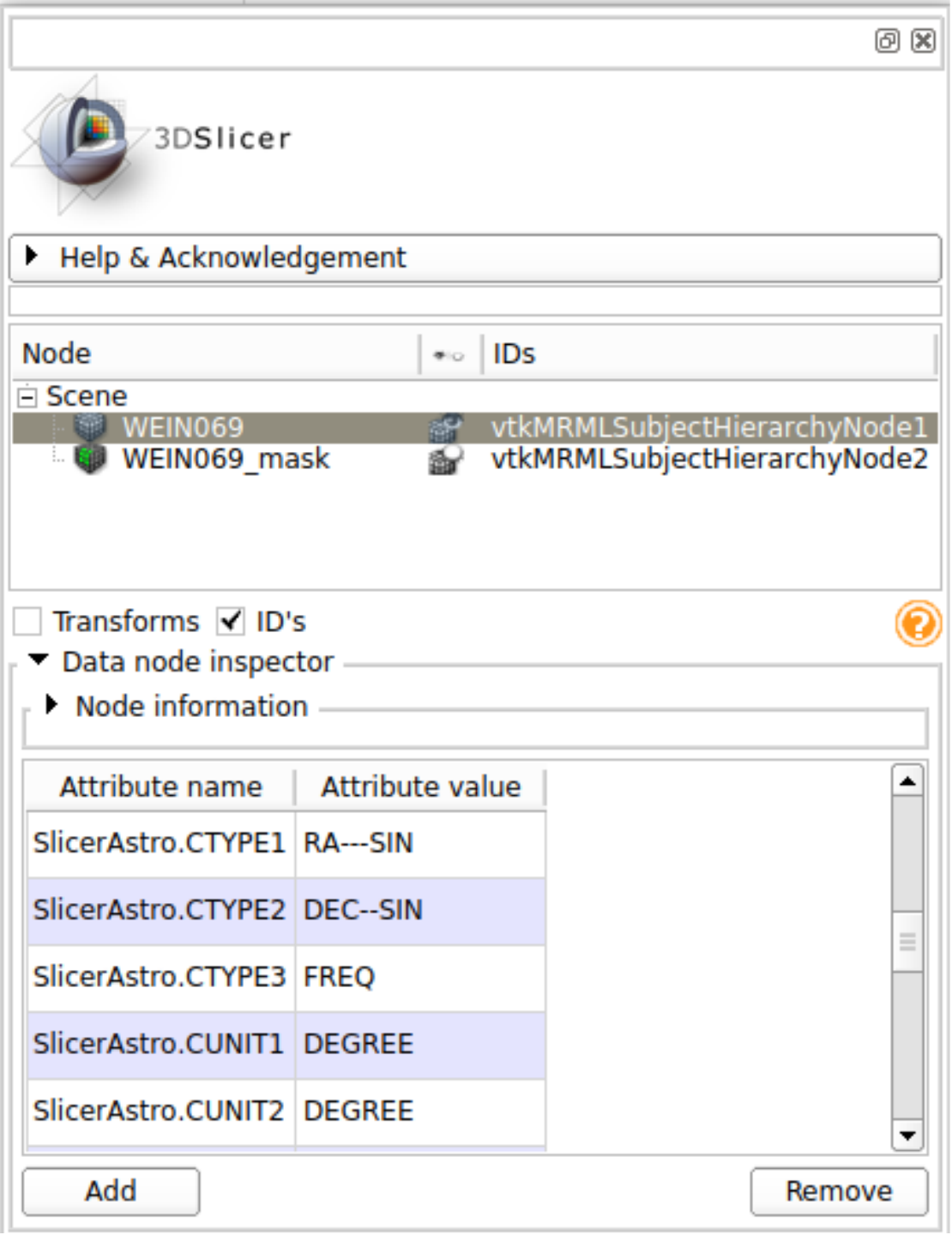}
\caption{The interface widgets of the $\tt{SubjectHierarchy}$ module. 
In the top panel, the interface includes the widgets for selecting
MRML nodes representing the data-cubes.
In the bottom panel, the interface includes a tool to inspect 
and modify the FITS keywords.
}
\label{SlicerAstroFig0}
\end{figure}

\begin{sidewaysfigure*}
\centering
\includegraphics[width=1.\textwidth]{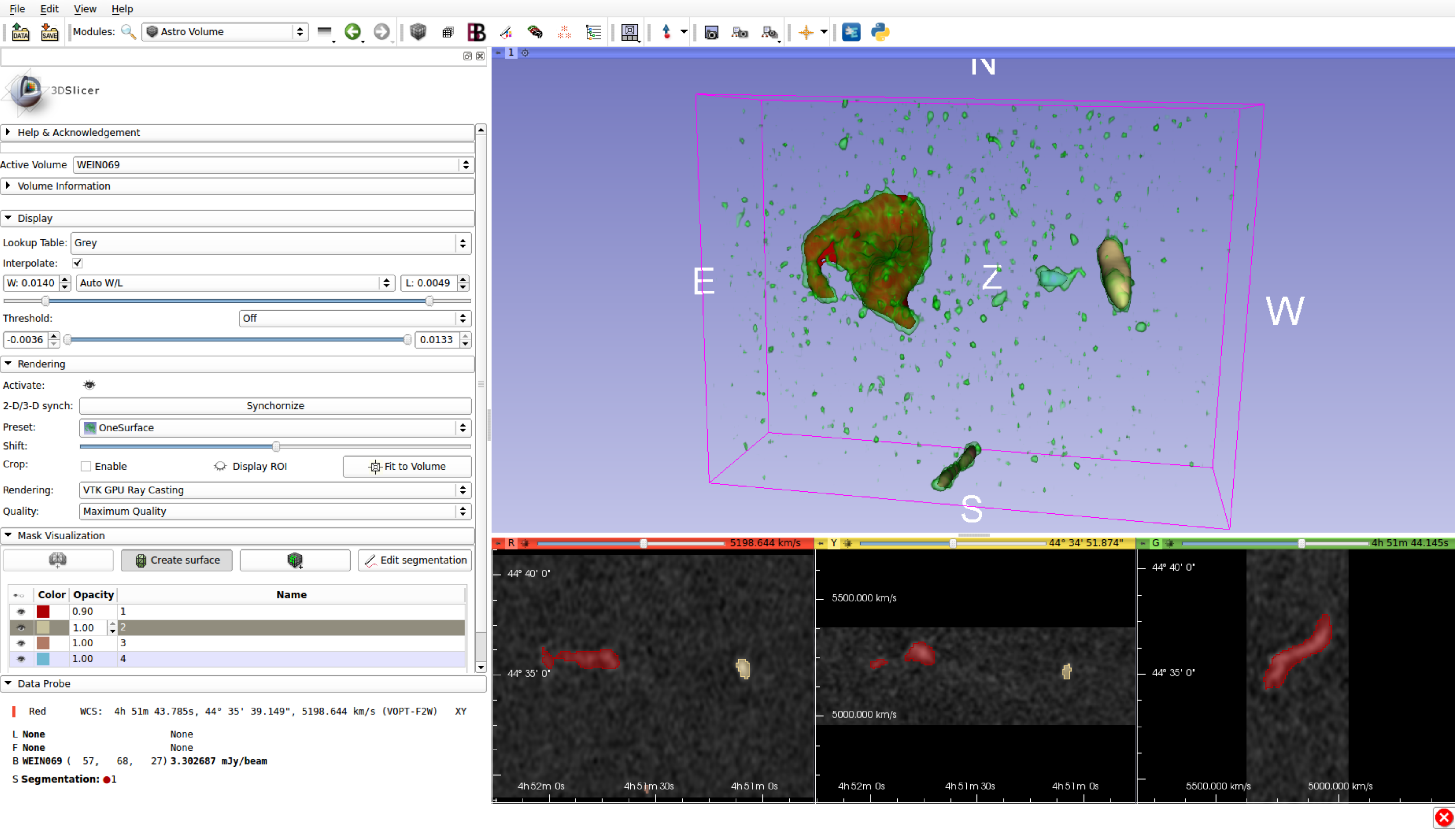}
\caption{Visualization of
the H\,{\scriptsize I} emission in and around WEIN069 \citep{Mpati} 
and its mask, generated with $\tt{SoFiA}$, in $\tt{SlicerAstro}$.
The data-cube contains three sources, i.e., WEIN069 and two companion galaxies,
a tidal tail and a very faint filament that connects 
two galaxies. In the 3-D view the data are rendered in green and highlighted 
at an intensity level equal to 3 times the root mean square ($rms$) noise. The colored segmentations
represent the mask and each color refers to a specific source ID as shown
in the table widget in the left panel.
The left panel includes also interface widgets to control 
the 2-D and 3-D color transfer functions and
a data probe window. 
Quantitative information such as WCS coordinates are shown both in the data 
probe window and along the axes in the 2-D views. The white labels 
in the 3-D view represent the four cardinal directions (\textit{N}, \textit{S}, 
\textit{E}, \textit{W}) and the line-of-sight direction (representing frequency/wavelength or velocity/redshift, z, hence the symbol \textit{Z}).
}
\label{SlicerAstroFig1}
\end{sidewaysfigure*}

In Fig.~\ref{SlicerAstroFig1}, we show the implementation of 
the $\tt{3DSlicer}$ and $\tt{SlicerAstro}$ interface.
On the top, the main menu shows several options for loading and writing files 
(including FITS files) and for editing the $\tt{3DSlicer}$ settings.
The data loaded from a FITS file are stored in a 
$\tt{vtkMRMLAstro}$-$\tt{VolumeNode}$ object.
The instantiated MRML nodes and their properties can be inspected in the 
$\tt{SubjectHierarchy}$ module (see Fig.~\ref{SlicerAstroFig0}). 
The output of source finder pipelines, that is,
\textit{object masks}, are loaded as 
$\tt{vtkMRMLAstroLabel}$-$\tt{MapVolumeNode}$ objects.
These masks are delivered as a data-cube  
where non-detected voxels in the original data-cube have a value of 0 and
detected voxels have an integer value corresponding to the ID of
the object they belong to. Due to the complex 3-D nature of the sources
\citep{Sancisi} and the noisy character of the data,
constructing a fully automated and reliable pipeline is not trivial 
\citep{Popping} and visualization can help in identifying or rejecting very faint signals \citep{Punzo2016}.
For example, in Fig.~\ref{SlicerAstroFig1}, $\tt{SlicerAstro}$ shows the visualization of 
the H\,{\small I} emission in and around WEIN069
and its mask, generated with $\tt{SoFiA}$ \citep{sofiaCode}.
The data-cube contains three sources, WEIN069 and two companions,
each identified as a separate source with its own mask. In addition 
there is a tidal tail and a very faint filament that is
connecting two of the galaxies. 

The left panel in Fig.~\ref{SlicerAstroFig1}, includes the widgets
for changing the the 2-D and 3-D color transfer functions for 
$\tt{vtkMR}$-$\tt{MLAstroVolumeNode}$ objects.
In the case of $\tt{vtkMRMLAstro}$-$\tt{LabelMapVolumeNode}$ objects, 
volume rendering is not available, but
it is possible to use the $\tt{MaskVisualization}$ widget
to convert the $\tt{vtkMRMLAstroLabelMapVolumeNode}$ object 
to a $\tt{vtkMRMLSegmentationNode}$ object.  
The $\tt{vtkMRML}$-$\tt{SegmentationNode}$ class is
a core class of $\tt{3DSlicer}$ that handles the 
display of data segmentation both in the 2-D and 3-D views, 
as shown in Fig.~\ref{SlicerAstroFig1},
and they can be overlaid on the data of a 
$\tt{vtkMRMLAstroVolumeNode}$ object. 
The segmentation objects can also be interactively 
modified (see Section~\ref{masking} for more information) 
and can be exported for 3-D printing (or imported in $\tt{Blender}$)
by saving them in the STL file format.

Moreover, the layout includes interface widgets to control the display
properties (e.g., user interaction to rotate the 3-D view),
a window displaying the 3-D World Coordinate and data values of the
position of a data probe in the linked 2-D views. 
The 2-D views also have quantitative World Coordinate axes.

Finally, the MRML infrastructure allows the user to save the session as a \textit{scene}. 
Reloading such a \textit{scene} restores the session. One can also 
share interesting visualizations with colleagues using the $\tt{Datastore}$
module. This module saves a bundle with all the necessary information
(the data, the visualization views, screen-shots and
text comments) on the Kitware servers. Other users can download these bundles.

\subsection{Rendering and user interactions}\label{rendering}
In $\tt{3DSlicer}$ the visualization representations are rendered with the
Visualization Toolkit, $\tt{VTK}$ \citep{Schroeder}.
In $\tt{SlicerAstro}$ the data are rendered in 3-D with the $\tt{VTK}$ implementation of the 
ray casting algorithm, a direct volume rendering method \citep{Roth}. 
Ray casting offers very high-quality results (i.e., free of artifacts),
but it is computationally expensive. On the other hand, ray casting is a massively 
parallel algorithm. On modern desktops the $\tt{VTK}$ GPU implementation
offers interactive rendering with a high ($> 5$) frame rate (FPS) for 
data-cubes not exceeding $10^9$ voxels.  
The use of such high quality rendering is mandatory in our case. 
In fact, other methods can produce many rendering artifacts in the 
noisy regions of an H\,{\small I} data-cube. 
In particular, indirect volume rendering
techniques are very ineffective at signal-to-noise-ratio $\lesssim 2$, because they have to fit 
geometries to very noisy data that do not have well-defined closed borders \citep{Punzo2015}. 

In $\tt{SlicerAstro}$, the masks are visualized as segmentations (i.e., $\tt{3DSlicer}$ 
renders them with indirect volume rendering), because they are supposed to be noise-free 
by definition. 

$\tt{3DSlicer}$ offers several 2-D/3-D linked navigation and interaction tools such as 
\textit{crosshair}, \textit{fiducials}, \textit{region of interest} (ROI), \textit{ruler}
and slice views linked with 3-D views 
\citep[for more information, we refer to][ and the $\tt{3DSlicer}$ online documentation\footnote{ 
\url{https://www.slicer.org/wiki/Documentation/Nightly}}]{Slicer}. All these features are
extremely useful for navigating and probing the data. However, the 3-D visualization paradigm 
used in $\tt{3DSlicer}$ and $\tt{SlicerAstro}$ is limited by 
the use of 2-D input and output hardware such as a standard monitor and mouse. 
An obvious limitation in 3-D is that it is not 
straightforward to select features 
or pick positions (i.e., voxels) in the 3-D space in an intuitive manner. 
Complementary visualization in 
2-D (linked to the 3-D one) can partially address these deficiencies.
 
In $\tt{3DSlicer}$ all the modules are accessible at run-time from the
$\tt{Python}$ console ($\tt{Python}$ version 2.7.11 
is bundled and delivered together with the $\tt{3DSlicer}$ binaries). 
Note, however, that of the packages often used in astronomy
only $\tt{numpy}$ is part of this bundle. This allows additional flexibility for
user-customized visualization and analysis tasks using all $\tt{3DSlicer}$ and 
$\tt{SlicerAstro}$ capabilities.
The $\tt{Python}$ console and automated $\tt{Python}$ 
scripts are a very powerful 
tool for interacting with the data itself. Some examples are:
accessing the array containing the data,
modifying the data and calculating statistics in a region of interest. Moreover, 
the MRML objects store everything that the user visualizes 
and changes in the interface. This allows the user to perform the same actions
by using $\tt{Python}$ scripts. An example for applying smoothing to a data-cube, performing 
the rendering, and saving the result as a video
is shown in the appendix, Section~\ref{Appendix}.
For example, this framework is extremely useful 
for creating screenshots and videos for a large number of sources. 
For more information, we also refer to the online
documentation\footnote{\url{https://github.com/Punzo/SlicerAstro/wiki}}.

\section{Interactive filtering}\label{filtering}

Future blind H\,{\small I} surveys will detect a 
large variety of galaxies with additional 
complex features such as tails, extra-planar gas, and
filaments. These faint structures can be found in nearby, well 
resolved galaxies and groups of marginally resolved galaxies.
They have a very low signal-to-noise
ratio ($\sim 1$), but are extended over many pixels.
Efficiently separating such signals from the noise
is not straightforward \citep{Punzo2016}.
Moreover, in the case of Apertif \citep{Apertif3} and ASKAP \citep{askap}, 
it is estimated that tens of such sub-cubes will be 
collected weekly \citep{Punzo2015}.
This is a large volume of data, and a coupling between
the filtering algorithms and 3-D visualization can enhance
the inspection process of large numbers of galaxies and masks 
provided by source finder algorithms.

\begin{figure}[!ht]
\centering
\includegraphics[width=0.48\textwidth]{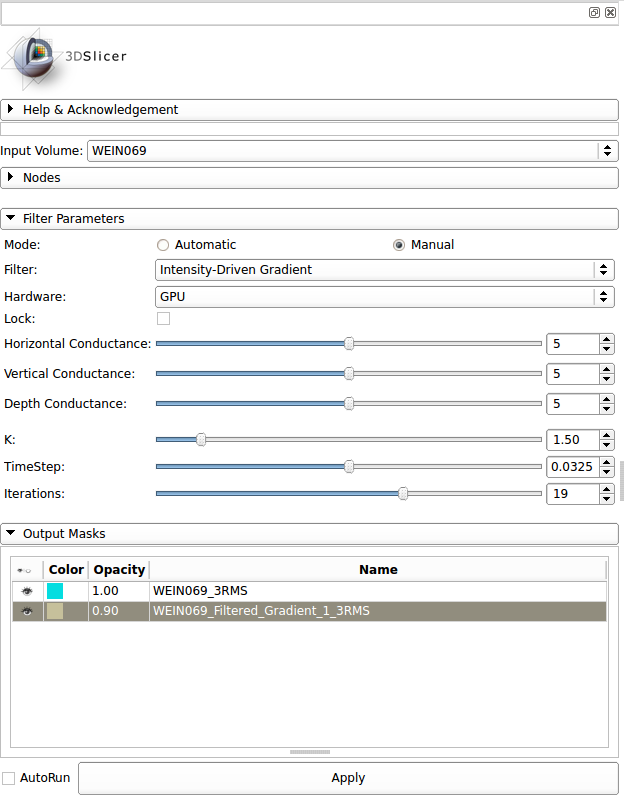}
\caption{The interface widgets of the $\tt{AstroSmoothing}$ module. 
The interface includes a widget for changing the input parameters for the smoothing and 
a table showing the output segmentations generated after the smoothing process.
}
\label{SlicerAstroFig2Widgets}
\end{figure} 

\begin{figure*}[!ht]
\centering
\includegraphics[width=1.\textwidth]{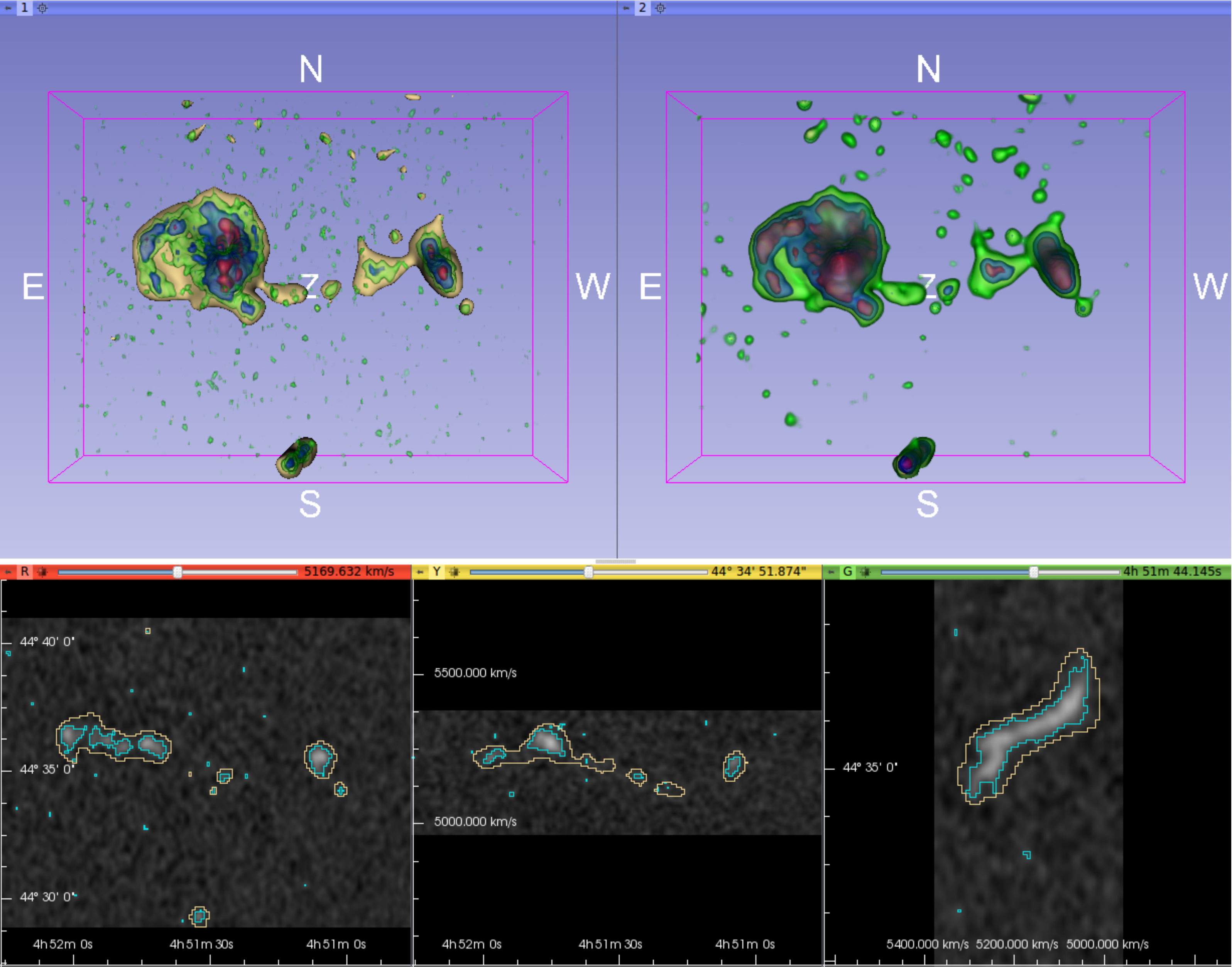}
\caption{A comparative layout of the output generated by the $\tt{AstroSmoothing}$ module is shown.
In Fig.~\ref{SlicerAstroFig2Widgets}, we show the interface of the 
$\tt{AstroSmoothing}$ module. This includes the widgets
for changing the input (such as the filter choice, the computational 
hardware and the smoothing parameters) and visualizing 
the output segmentation objects generated by the smoothing process.
The layout is composed of two 3-D views and three 2-D views. In the left 3-D view and
the 2-D views the data are shown. In the right 3-D view the filtered version of the data is shown.
The data are rendered in different colors that highlight the data 
at different intensity levels: green, blue and red correspond to 3, 7 and 15 times the $rms$ noise respectively. The colored segmentations
represent masks automatically calculated by the filtering algorithm.
The light blue and yellow segmentations (visualized as contour plots in the 2-D views) are
a 3 $rms$ thresholding of the input data and the filtered data, respectively.
}
\label{SlicerAstroFig2}
\end{figure*}

Three filters are currently available in the $\tt{AstroSmoo}$-$\tt{thing}$ module:
\begin{enumerate}[1)]
\item \textit{Box} (or \textit{mean}) filter: it replaces
each pixel value in the volume with the mean value of its neighbors
including the value of the pixel itself. Both isotropic (where
the 3-D kernel has the same dimensions along all the three axes) and
anisotropic implementations are available.
\item \textit{Gaussian}: it applies a 3-D convolution operator to the volume. 
It preserves the shape of the objects better than the box filter,
but is computationally more expensive.  
Both isotropic and anisotropic (the kernel can have 
different dimensions along the 3 axes and it can be 
rotated) implementations are available.
\item \textit{Intensity-Driven Gradient}: it uses an adaptive diffusion 
process (i.e., operates on the differences between 
neighboring pixels, rather than on the pixel values directly).
High signal-to-noise regions are unaffected, but low signal-to-noise, 
extended, regions are enhanced.
\end{enumerate}

These algorithms are available in $\tt{SlicerAstro}$ as parallelized implementations 
on both CPU and GPU hardware, offering interactive performance when processing data-cubes
of dimensions up to $10^7$ voxels and very fast
performance ($< 3.5$ sec) for larger ones (up to $10^8$ voxels).
The intensity-driven gradient filter, due to 
its adaptive characteristics, is the optimal choice for H\,{\small I} data \citep{Punzo2016}.
Therefore, it is the default method when the automatic mode has been chosen. This algorithm preserves the detailed
structure of the signal with high signal-to-noise ratio ($> 3$)
at the highest resolution, while smoothing only the faint part of the signal 
(signal-to-noise ratio $< 3$). For more information regarding the filters and their
performance, default parameters, advantages and disadvantages, we refer to \cite{Punzo2016}. 

After running the smoothing process, $\tt{SlicerAstro}$ displays automatically a
comparative layout composed of two 3-D views, one of the original data 
(top left panel) and one of the filtered data (top right panel), 
and three 2-D views of the original data (lower three panels) for 
the inspection of the data,
as shown in Fig.~\ref{SlicerAstroFig2}. In this particular case, the  
3-D visualization of the filtered data highlights immediately the presence of the faint filament
between two galaxies that was hardly visible in the original version of the data. 
Moreover, the coupling between 
3-D visualization and interactive filtering enables a user to manually and iteratively 
search the best smoothing parameters for maximally enhancing the local signal-to-noise ratio
of the very faint component. 

We will show in the next section how any segmentations generated by the
smoothing module (or converted from loaded 
masks as shown in Section~\ref{framework}) can be interactively modified 
in the $\tt{SegmentationEditor}$ module of $\tt{3DSlicer}$.

\section{Interactive 3-D masking}\label{masking}
Twenty years ago, \cite{Norris} pointed out that the main challenge for visualizing
astronomical data in 3-D was to develop a 3-D visualization tool with interactive capabilities 
for data inspection and with interactive and quantitative analysis
capabilities.
Nowadays, 3-D interactive visualization is achievable thanks to the use of massively
parallel hardware such as GPUs (see Section~\ref{rendering}). 
On the other hand, volumetric data interaction tools 
(e.g., picking a voxel or selecting a region
of interest in 3-D) are necessary for performing data analysis in a 
3-D environment.  

An optimized 3-D selection technique, based on 2-D input/output hardware, 
is still a partially open-problem, not only in 
astronomy, but also in medical visualization and computer science. 
Moreover, the optimal selection technique
highly depends on the specifications of the use case.
Our requirements for a 3-D selection tool
are interactivity and a minimal number of user-operations
for achieving the selection (i.e., user-friendliness).   
For a review of the state-of-the-art 3-D selection
algorithms we refer to \cite{Yu1} and \cite{Yu2}.

For our application, we opt for the $\tt{CloudLasso}$ technique \citep{Yu1}. 
The $\tt{CloudLasso}$, operated on grid data, is based on the application of the
Marching Cubes (MC) algorithm \citep{Wyvill, Lorensen} for the identification 
of regions of voxels with signal inside a user-drawn lasso; i.e., $\tt{CloudLasso}$ is
a lasso-constrained Marching Cubes method. The $\tt{CloudLasso}$ method allows us to spatially 
select structures with high signal-to-noise ratio ($> 3$) within a lasso region. 
Even if disjoint structures lie visually behind one another, 
they can be all selected without including the noisy regions in between. 
For operating the intended selection, a
threshold has to be chosen. The $\tt{CloudLasso}$ algorithm,
therefore, comprises the following two steps:

\begin{figure}[!ht]
\centering
\includegraphics[width=0.48\textwidth]{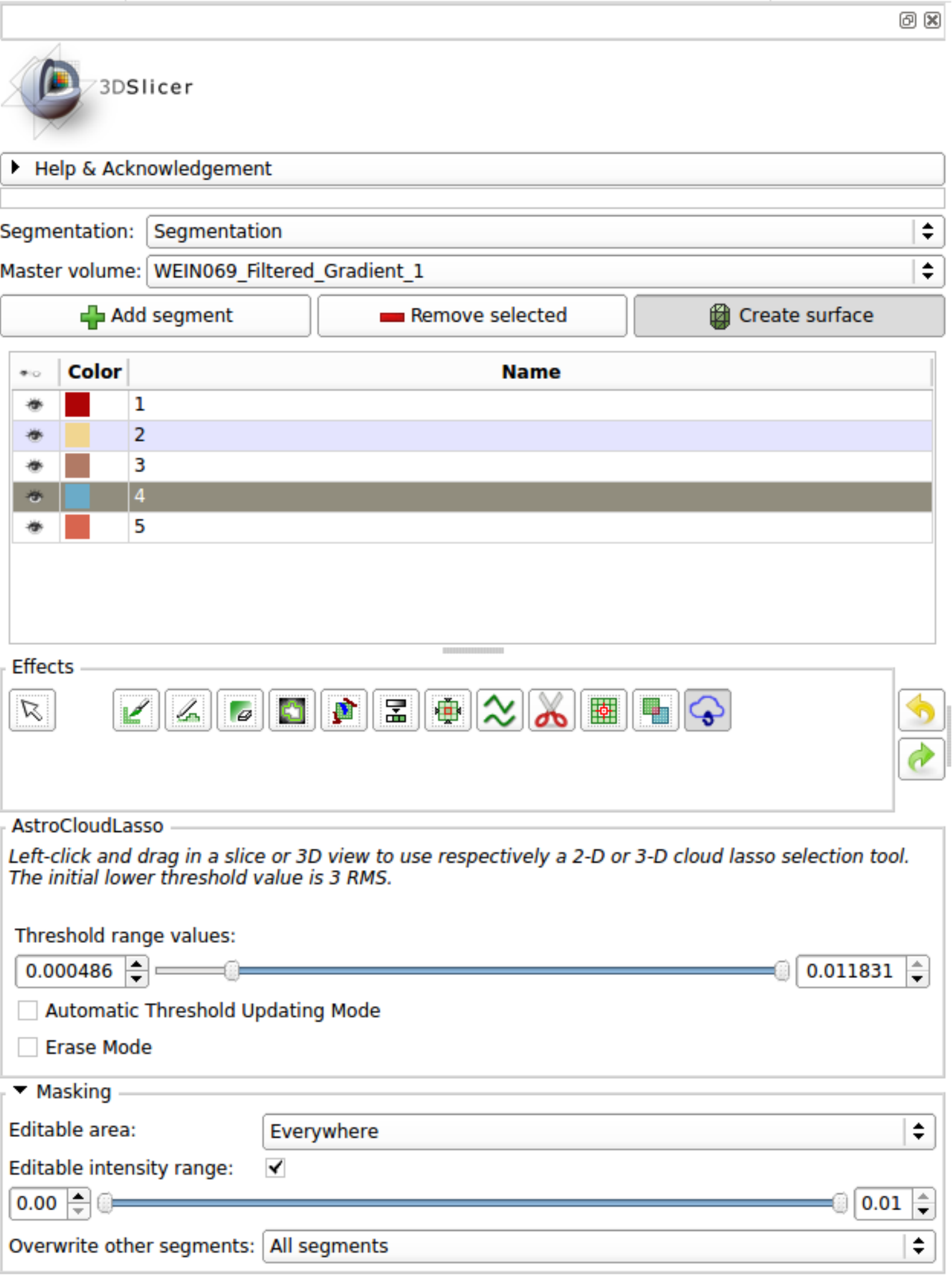}
\caption{The interface widgets of the $\tt{SegmentationEditor}$ module and
$\tt{AstroCloudLasso}$ segmentation effects. 
The interface includes widgets for selecting
the segment to modify, the segmentation editor effect and the parameters relative to
the chosen effect.
}
\label{SlicerAstroFig3Widgets}
\end{figure}

\begin{enumerate}[A)]
\item Volume selection: the subset of the volume where the intensities of the voxels
exceed a threshold is computed using Marching Cubes.
\item Threshold tuning: interactive adjustment of the intensity threshold.
\end{enumerate}

The selection operated by the $\tt{CloudLasso}$ algorithm
is highly dependent on the user interactions. In fact, the user has to choose
the orientation of the camera which gives the best view of the data, 
perform the 2-D selection on the screen and select the optimal threshold. 
Therefore, the user experience and knowledge of the data are of primary importance
in the $\tt{CloudLasso}$ selection. The technique allows the user
to perform a refinement of the selection interactively 
by tuning the threshold or through Boolean Operations.

\begin{figure}[!ht]
\centering
\includegraphics[width=0.395\textwidth]{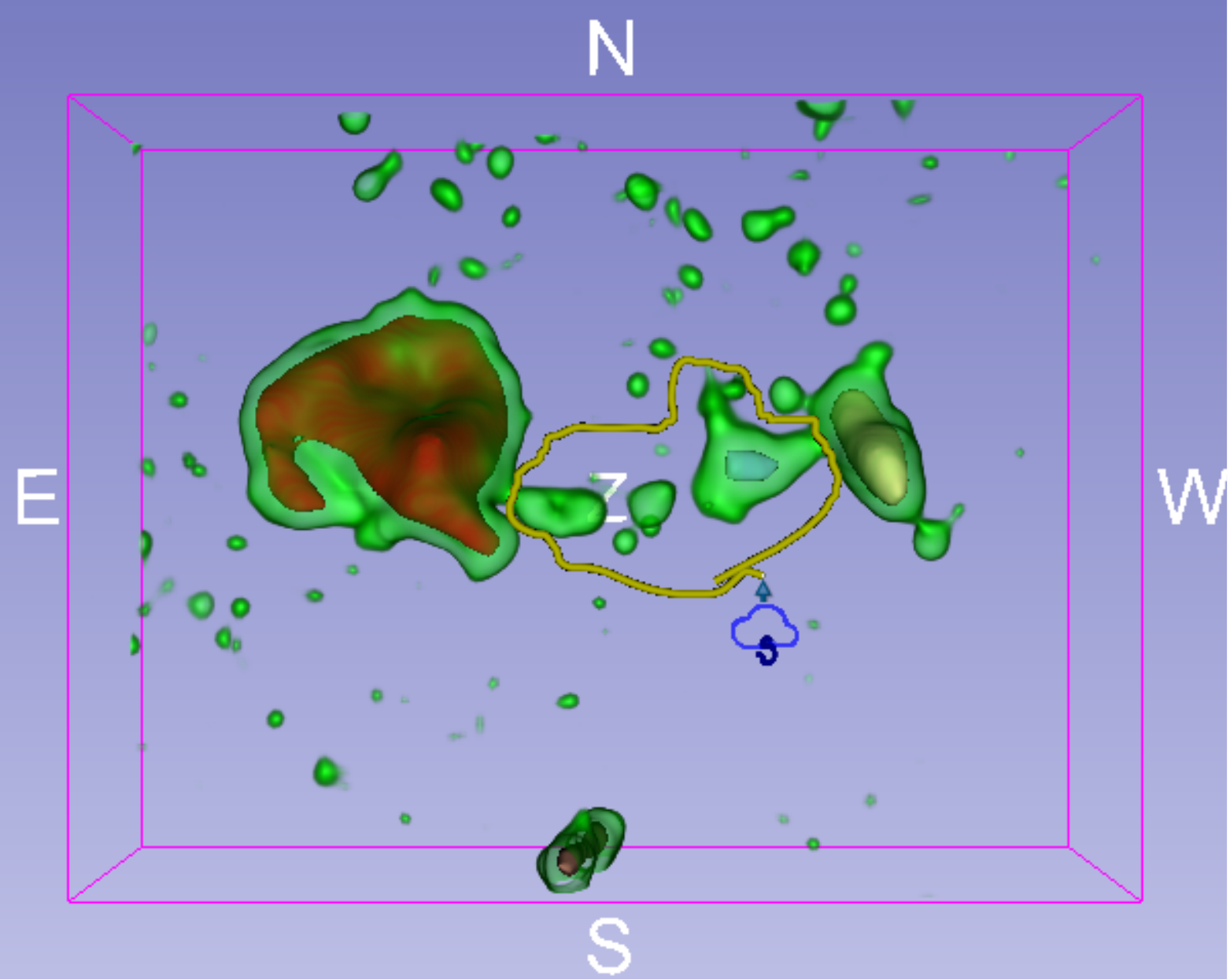}
\includegraphics[width=0.395\textwidth]{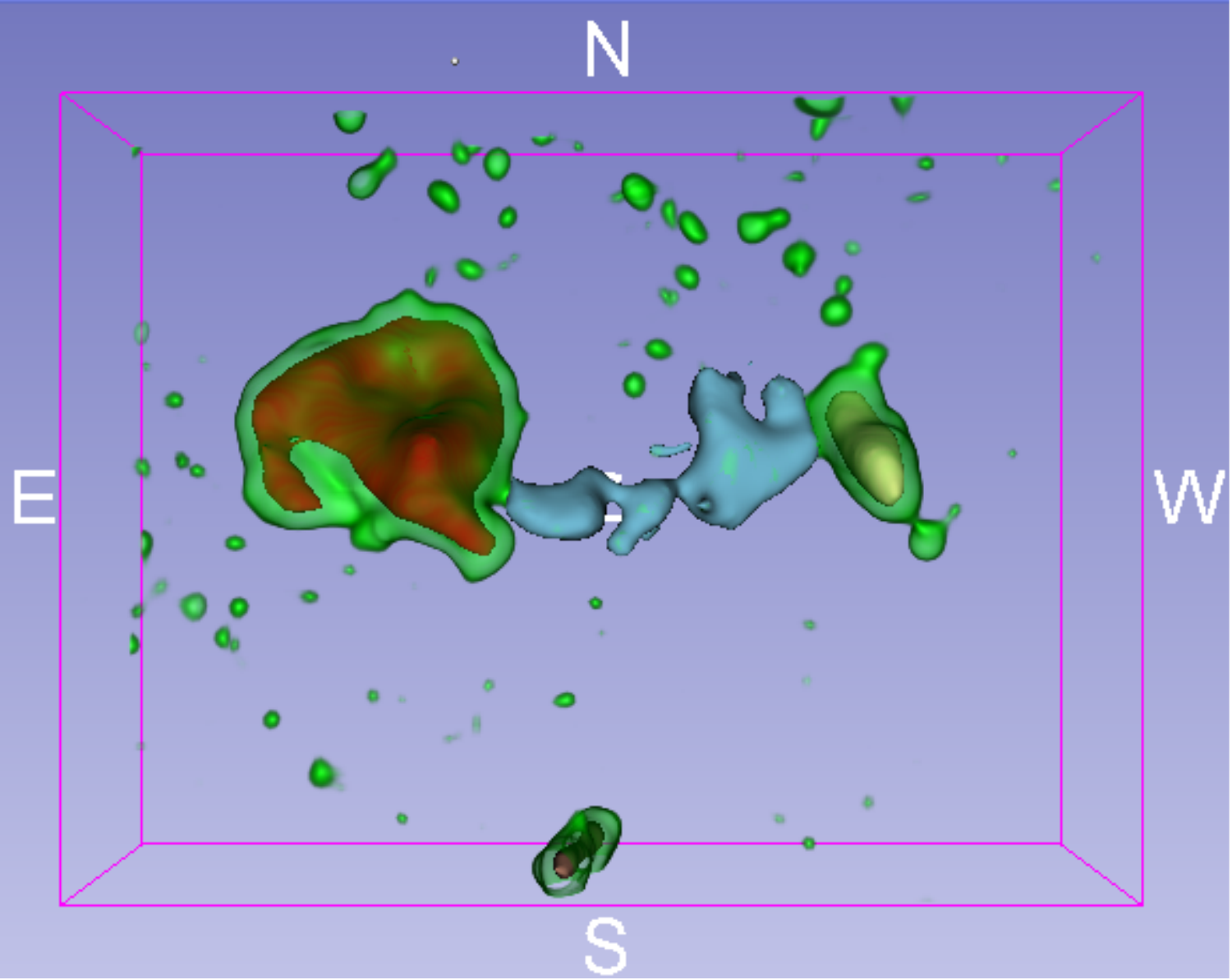}
\includegraphics[width=0.395\textwidth]{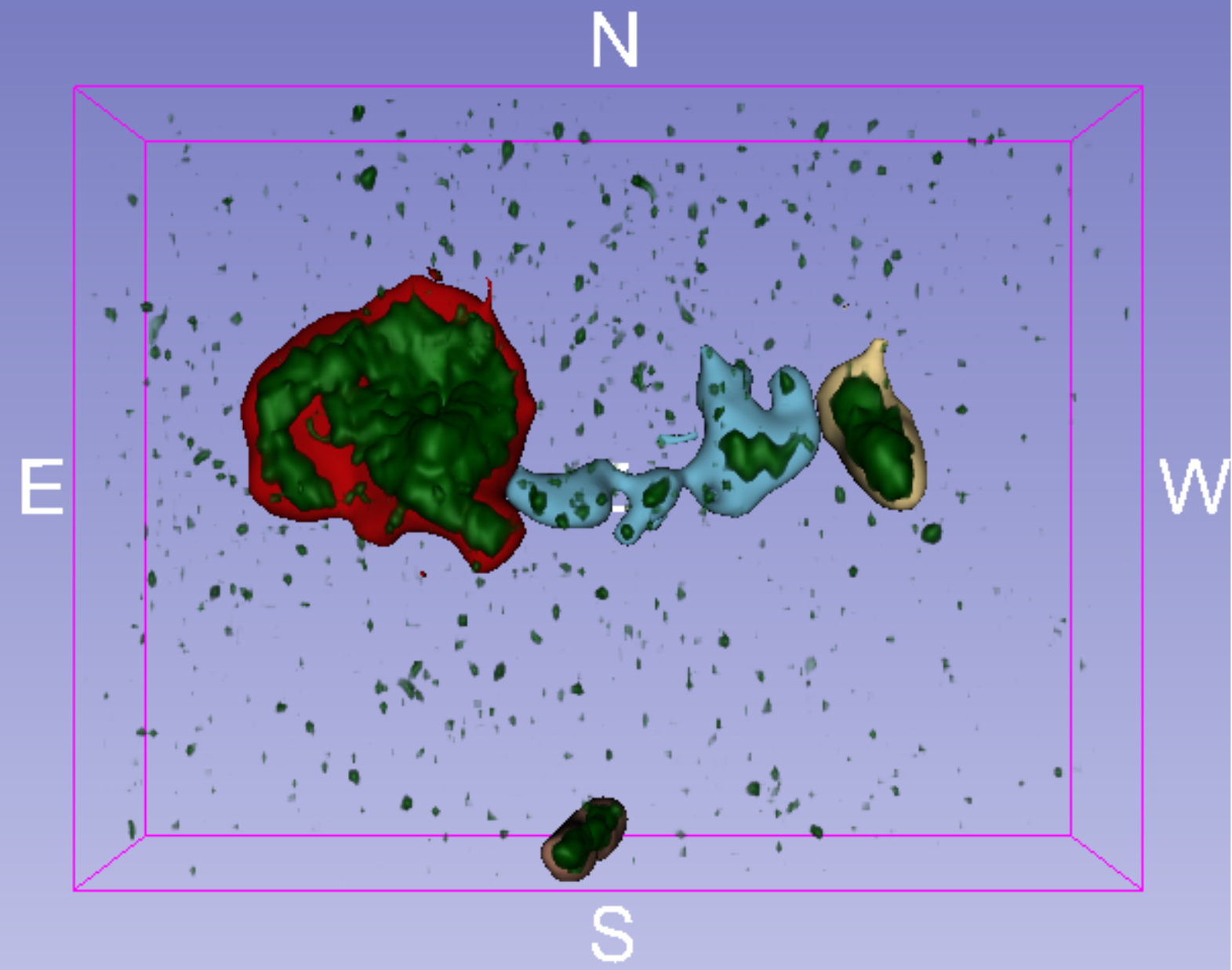}
\caption{Usage of the $\tt{AstroCloudLasso}$
segmentation editor effect in 3-D.
A smoothed version of the WEIN069 data is rendered in green
in the top and middle panels. In the bottom panel, the 
original version of WEIN069 data is rendered. 
The three renderings highlight the data 
at the intensity level equal 3 times the $rms$. 
In the top panel, the colored segmentations
represent the mask shown in Fig.~\ref{SlicerAstroFig1}.
In order to visualize clearly both the data and the mask, 
the data are rendered with a higher opacity in the bottom
panel compared to the upper panels.
Similarly, the opacity of the segmentations is decreased. 
The $\tt{AstroCloudLasso}$ selection tool
is visualized as a yellow tube
drawn by the user with the 2-D cursor indicated by the blue cloud. 
This tool computes a selection in 3-D space from the 2-D user-selection.
It builds a closed surface at the value of the intensity level specified 
in the settings widget (Fig.~\ref{SlicerAstroFig3Widgets}) and
visualizes the modified segment as shown in the middle panel.
In the bottom panel, we show all the modified segments.}
\label{SlicerAstroFig3}
\end{figure}

Although the threshold tuning step can be improved
or replaced by more complex techniques to identify and classify 
the signal in the selection, the $\tt{CloudLasso}$ technique 
is the most reliable choice in our case, because it leaves any 
classification to the user (leveraging his/her knowledge 
about the data). For example, 
connectivity operators \citep{Heijmans} can be applied after the thresholding
to distinguish the various islands of signal and to label them with IDs.
Moreover, $\tt{MAX}$-$\tt{TREE}$ algorithms \citep{moschini2014} can automatically provide 
a tree classification of the data. Finally, more 
advanced selection techniques can be employed
\citep[e.g., $\tt{Cast}$ selections][]{Yu2}. The common element in these techniques is 
the idea to classify (in different ways) the information in the data. 
However, due to the very noisy nature 
of H\,{\small I} data, separating the H\,{\small I} signal from the noise
is not trivial \citep{Punzo2016} and, therefore, it is quite challenging to 
build an automated algorithm to classify the data see also \citep[see also][]{Giese2016}.

In the $\tt{SegmentationEditor}$ module of $\tt{3DSlicer}$ we implemented
an $\tt{AstroCloudLasso}$ segmentation editing capability, 
optimized and specialized
for the selection of H\,{\small I} data. A segmentation editor is a $\tt{3DSlicer}$ 
tool that enables user interaction with the data and creation/modification of segmentations 
both in the 2-D and 3-D views.
In Fig.~\ref{SlicerAstroFig3Widgets} we show the interface widgets 
of the $\tt{AstroCloudLasso}$ segmentation editor. 
The default value of the threshold is set to 3 times the $rms$ 
value of the data-cube under study. 
In Fig.~\ref{SlicerAstroFig3}, we show how the selection procedure is performed
in a 3-D view of $\tt{3DSlicer}$ and
the results for each segment are shown
(i.e., we repeated the selection procedure four times).
The tool can also perform 2-D selections (on the 2-D views),
it can erase the segment under the 
selection (both in the 2-D and 3-D views) if the erase mode has been enabled, and
it can interactively adjust the selection of the intensity threshold if 
the automatic updating mode has been enabled.

The $\tt{AstroCloudLasso}$ segmentation editor effect can be used for two applications: 
\begin{enumerate}[A)]
\item interactively modify a mask as shown in Fig.~\ref{SlicerAstroFig3} 
(note that $\tt{SlicerAstro}$ can save the new mask as a FITS file). 
This framework can be used as a modification tool 
of the masks generated from source finder pipelines. 
\item selecting regions of interest for further analysis.
\end{enumerate}

In the next section, we will apply the segmentation 
as a selection for operating tilted-ring modeling in the region of interest. 

\section{Interactive modeling}\label{modeling}

In the case of H\,{\small I} in galaxies
one can extract additional information from fitting the 
observations with a so called \textit{tilted-ring}
model \citep{Warner}. Such a model describes
the observed H\,{\small I} distribution of the galaxy as a set 
of concentric, inclined, and rotating rings. 
Each ring is characterized by the following parameters: the 
center of the spatial coordinates, and the systemic velocity, rotation 
velocity, velocity dispersion, inclination, position angle as a 
function of the galactocentric radius. A model is specified by a set
of ring (radially varying) parameters plus a set of global parameters 
(e.g., ring width).

To compute a model the rings are populated with an ensemble of H\,{\small I} 
clouds using a Monte Carlo method. The cloud ensembles are integrated along
each line-of-sight in the data-cube and convolved with a 3D-Gaussian 
representing the properties of the observing beam and the resolution in the frequency domain.

A tilted-ring model is necessarily an oversimplification of 
the H\,{\small I}  distribution inside galaxies.
When the orbits are significantly non-circular, for example
in the presence of a bar \citep{Bosma}, the tilted-ring model will not
be able to represent the data accurately. Furthermore, there is a degree
of degeneracy between some of the ring parameters
(e.g., inclination, position angle and rotational velocity). 
In many cases, however, the tilted-ring model serves as
a good approximation and can provide a deeper understanding of the
kinematics and morphology of a galaxy, including asymmetries in surface
density and velocity, the presence of gas at anomalous velocities,
of extra-planar gas, of inflows 
and outflows, etc. It is for example rather easy 
to locate the presence of extra-planar gas once the symmetric and 
regularly rotating disk is modeled (see Section~\ref{UseCaseB}). 

It is, therefore, very useful to add
model fitting capabilities to a visual analytics tool for H\,{\small I} data. Such a capability  
enables an interactive comparison between the data and models so that 
the quality of the model can be assessed interactively. This is possible 
by embedding the model routine in the visualization interface. This 
will also enable interactive tuning of the model parameters using the visualization interface.

Modern 3-D tilted-ring modeling software 
can generate symmetric models that reproduce the data 
with a minimal user input and interaction \citep{Teodoro, Kamphuis}.
In the next section we will briefly review such software libraries. We 
will also describe the integration of one of them in $\tt{SlicerAstro}$ 
and show how it provides additional capabilities for the detection and analysis 
of subtle structures in the 3-D domain. Two use cases will be investigated
in Sections \ref{UseCaseA} and \ref{UseCaseB}: using the 
3-D selection tool (shown in Section~\ref{masking}) to perform the tilted-ring model
fitting only in a region of interest (i.e., excluding non-symmetric, non-regular, 
H\,{\small I} structures such as tidal tails) and using the symmetrical properties
of automated tilted-ring model fitting to locate extra-planar gas.

\subsection{Requirements}
Tilted-ring model fitting is rather complex. Therefore we chose to rely 
on an external state-of-the-art package rather than designing a new one. 
In order to be able to wrap an external model fitting package into
$\tt{SlicerAstro}$ the following requirements can be formulated:
\begin{enumerate}[I)]
\item 3-D model fitting capabilities;
\item automatic estimation of the initial parameters for the fitting;
\item parallelization on CPU and/or GPU for fast execution;
\item developed as a modern, modular source code, preferably  \CC.
\end{enumerate}

Currently, two software packages are available: \newline
$\tt{^{\rm3D}\,Barolo}$ \citep{bbarolo}, an automated procedure 
which fits tilted-ring models to H\,{\small I} data-cubes; and $\tt{Fat}$ 
\citep{FAT}, 
a similar package built on top of $\tt{TiRiFic}$ \citep{TiRiFic,Jozsa}.

\begin{table}[!ht]
  \centering
  \begin{tabular}{| c | c | c | c |}
    \hline
    Requirements   & $\tt{TiRiFic/Fat}$ & $\tt{^{\rm3D}\,Barolo}$ \\ \hline\hline
    
    I: fitting & & \\ 
    capabilities  & \ding{51} & 	\ding{51} 	  \\ \hline
    II: parameter & & \\ 
    estimation  &   \ding{51}	& 	\ding{51} 	  \\ \hline
    III: CPU/GPU & & \\ 
    parallelization &   \ding{55} & 	\ding{55} 	  \\ \hline
    IV: \CC & & \\ 
    development  &  \ding{55} &   \ding{51}     \\ \hline
  \end{tabular}
  \caption{Requirements for the model fitting external library,
  described in Section~\ref{modeling}.
  We compare two software packages: $\tt{TiRiFic/Fat}$ \citep{Jozsa,Kamphuis} and $\tt{^{\rm3D}\,Barolo}$ 
  \citep{Teodoro}.  }
  \label{ModelingRequirements}
\end{table}

As shown in Table \ref{ModelingRequirements}, $\tt{^{\rm3D}\,Barolo}$ is
currently our optimal choice because, being developed in \CC, it 
satisfies the fourth requirement.  
The fourth requirement ensures high performance, a rather simple and smooth
integration process, and long-term maintainability. 
On the contrary, $\tt{Fat}$ has been developed in $\tt{IDL}$ which
introduce with it several compiling and linking issues (also, 
the license of $\tt{IDL}$ is not compatible the open-source BSD license of $\tt{SlicerAstro}$).

Although both state-of-the-art packages 
lack of a concrete parallelization strategy, they still have 
sufficiently fast performance for fitting the data for sources in data volumes up to $10^6$ voxels and 
for generating a single model for sources representing a data volume 
of up to $10^8$ voxels (see Section~\ref{UseCaseA}). 
However, user interaction with the modeling routine in the
$\tt{AstroModeling}$ module will greatly benefit from a parallel 3-D
tilted-ring model fitting source code.

In the following use cases\footnote{In software and systems engineering, 
a use case is a list of actions or event steps, typically defining the 
interactions between a role (known in the Unified Modeling Language 
as an actor) and a system, to achieve a goal. The actor can be a 
human or other external system.} we will show how the $\tt{Astro}$-$\tt{Modeling}$ module,
exploiting 3-D interactive visualization and $\tt{^{\rm3D}\,Barolo}$, helps 
in the modeling and analysis of complex sources.

\subsection{Use Case A: analysis of sources with tidal tails}\label{UseCaseA}

Although $\tt{^{\rm3D}\,Barolo}$ is a powerful fitting routine, it is
designed to fit models of galaxies with a thin regularly rotating disk.
Therefore, $\tt{^{\rm3D}\,Barolo}$ (or
other current tilted-ring modeling algorithms) cannot recognize, for example, 
tidal tail structures and separate them from the central regularly 
rotating body of the galaxy.

In this section, we show how to use the $\tt{AstroModel}$-$\tt{ing}$ module
for the manual quality control of the models.  
This framework enhances the analysis of gravitational perturbed
galaxies such as WEIN069. In fact, the 3-D selection tool described
in Section~\ref{masking} can be used to select a region of interest 
for which $\tt{^{\rm3D}\,Barolo}$ provides the best results.
For example, in the case of WEIN069 the user can separate the two kinematic components,
i.e., the regularly rotating disk and the tidal tail, and perform the 
calculations only on the central disk. 

\begin{figure}[!ht]
\centering
\includegraphics[width=0.48\textwidth]{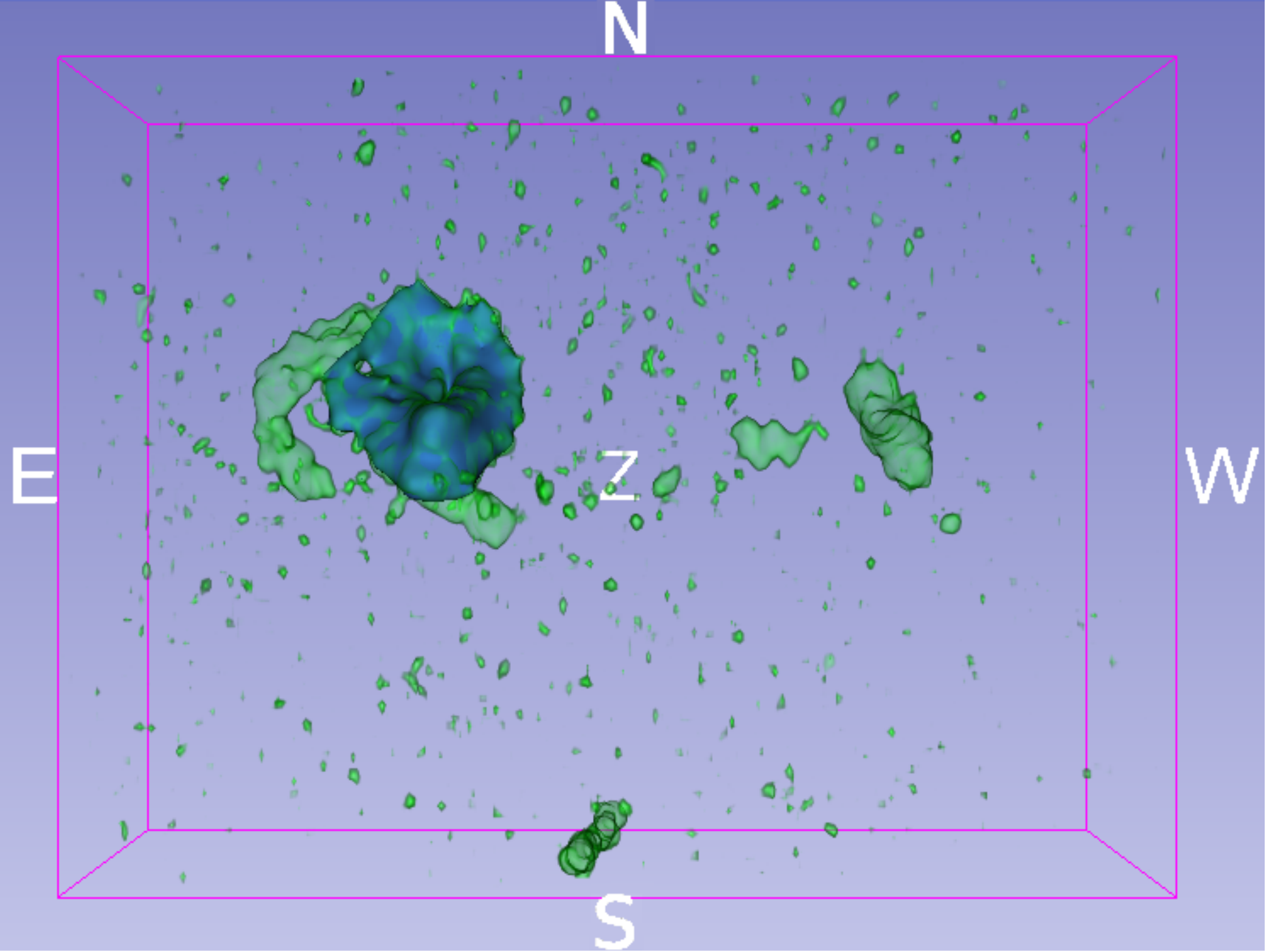}
\caption{3-D view of WEIN069 rendered in green. It highlights the data 
at the intensity level equal 3 times the $rms$. The blue
segmentation represents a 3-D selection (see Section~\ref{masking}).}
\label{SlicerAstroFig5}
\end{figure}

\begin{figure}[!ht]
\centering
\includegraphics[width=0.42\textwidth]{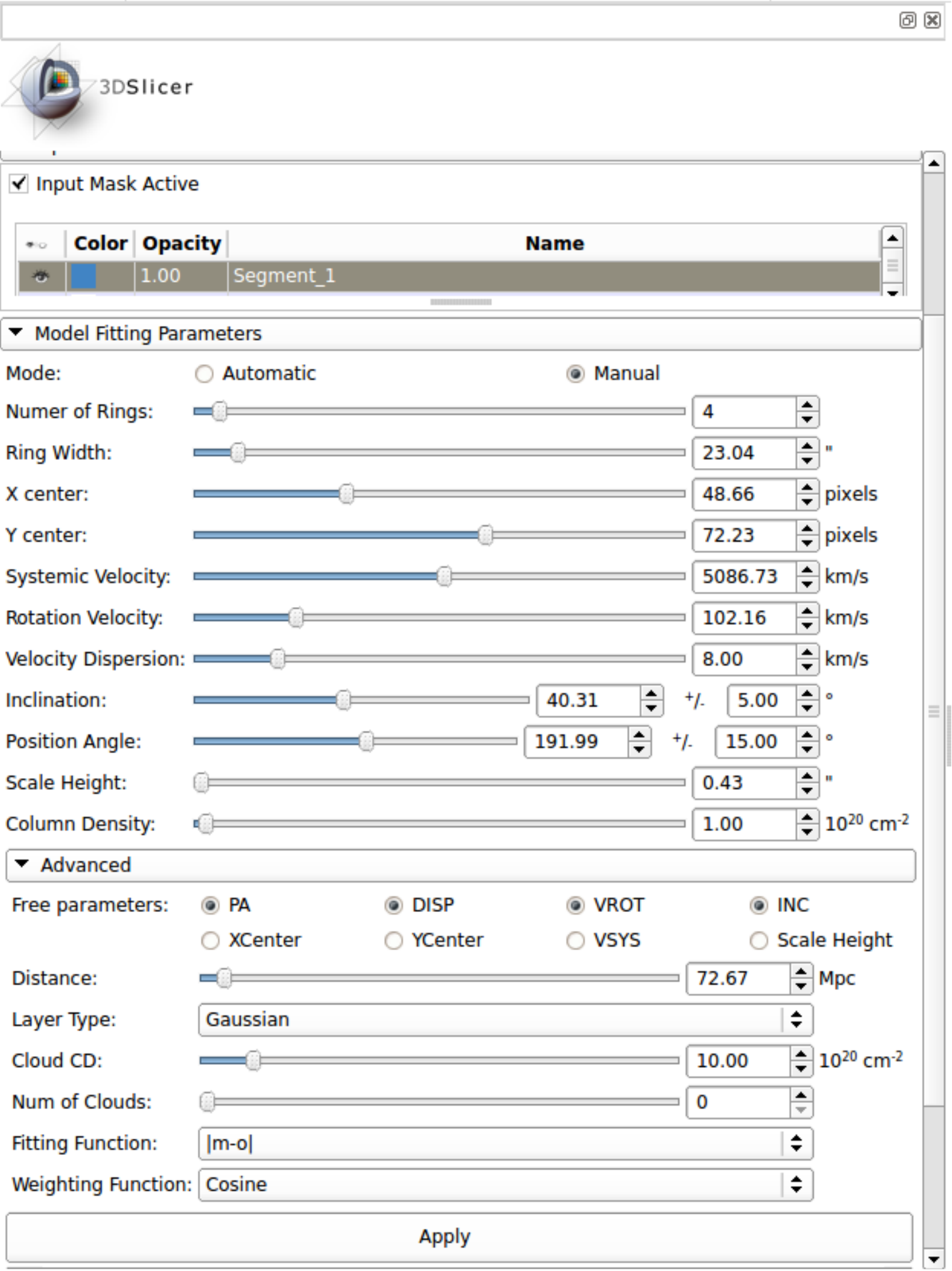}
\includegraphics[width=0.42\textwidth]{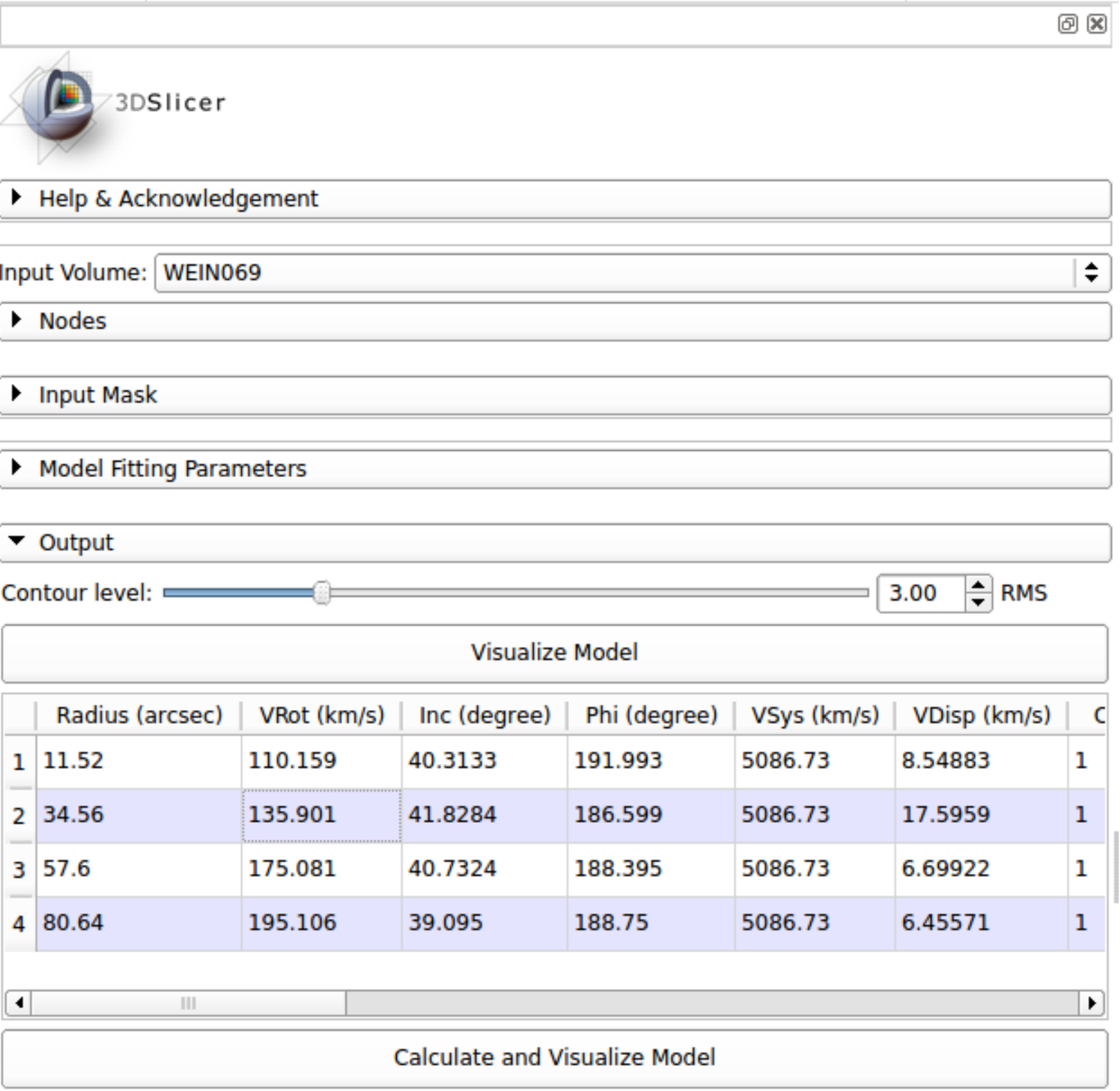}
\caption{The interface widgets of the $\tt{AstroModeling}$ module. 
In the top panel, the interface includes the widgets for selecting
a segment that will be used as mask for the modeling and the input parameters 
for the model fitting. In manual mode one can specify the fitting method 
(i.e.~the kind of residuals between the model and the data to be minimized) and the
weighting function, i.e.~the weights (as a function of angle from the minor axis) given 
to the residuals before fitting. These weights correct for the effect that the line of
sight component of the circular velocity in a rotating disk approaches zero when
approaching the minor axis. 
\citep[for more information see][]{bbarolo}.
In the bottom panel, the interface includes a \textit{Contour level}
widget to choose the threshold value for the segmentation of the model,
an editable table with the parameters of the rings of the 
output model (see Fig.\ref{SlicerAstroFig6}), and 
push buttons for updating the model.}
\label{SlicerAstroFig5Widgets}
\end{figure}

\begin{figure*}[!ht]
\centering
\includegraphics[width=0.89\textwidth]{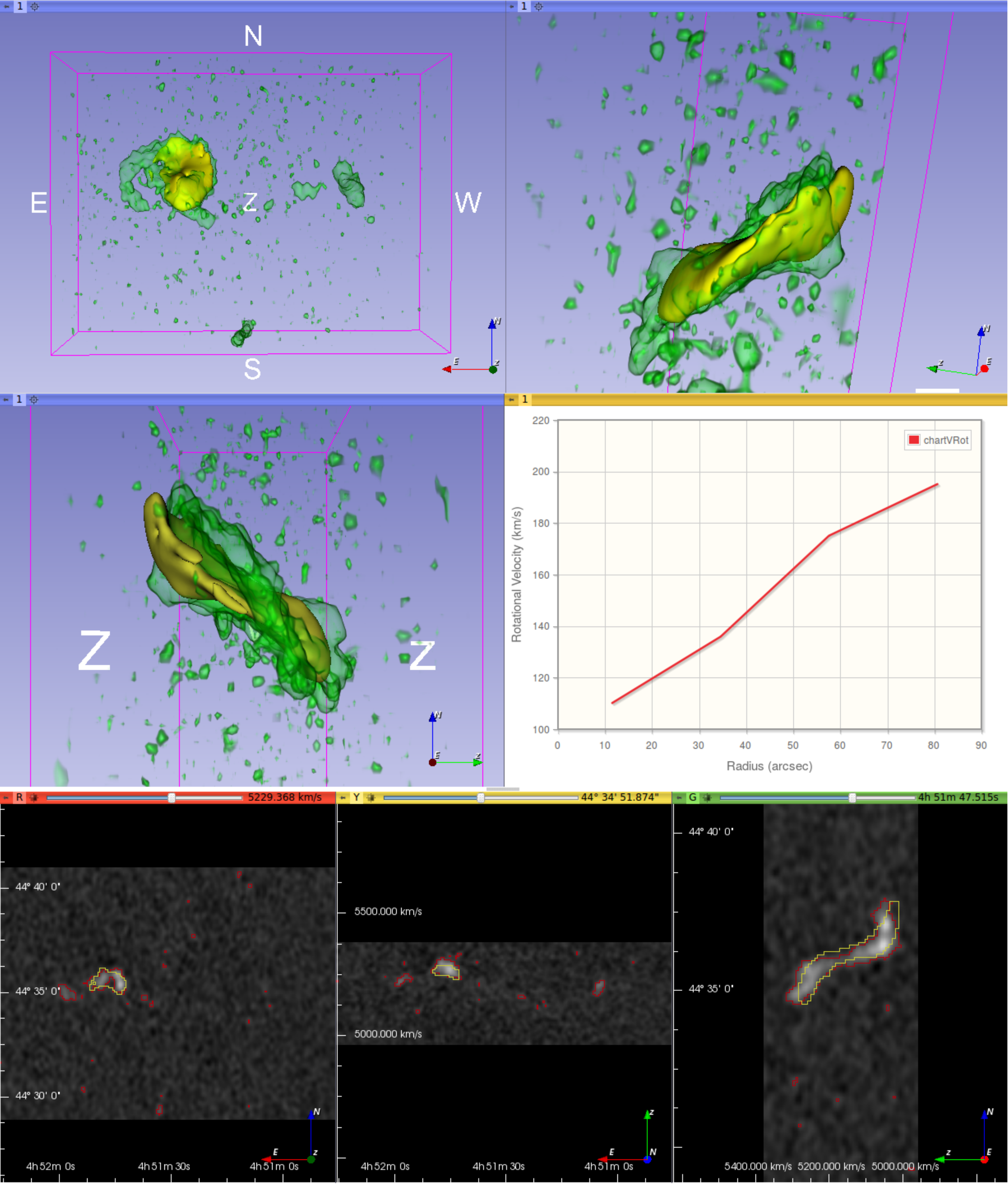}
\caption{Comparative layout of the output generated by 
the $\tt{AstroModeling}$ module.  
The layout is composed of three 3-D views, 
three 2-D views and a chart view. 
The WEIN069 data are shown in the 3-D view and
the 2-D views. In the 3-D views the data are rendered 
in green. Each 3-D view has a different camera position:
top-left, viewing direction along the velocity axis;
top-right, the origin of the camera
is in the center
of the data-cube and the view is parallel to 
the geometrical major axis of the galaxy; 
middle-left, the
viewing direction is along the RA axis. The white labels 
\textit{z} and \textit{Z} indicate the line of sight 
velocity (or redshift z) direction (i.e., 
increasing from \textit{z} to \textit{Z}). 
The green arrow in the lower right corner of the top 
right panel points to the plane indicated by the symbol \textit{z}.
The middle-right view has plotting capabilities and the different parameters
of the rings of the output model can be shown. The bottom views are slices of the data-cube:
bottom-left, XY; bottom-middle, XZ; bottom-right, ZY.
In the 2-D views the data are displayed with a grayscale color function.
The yellow segmentation (in the 3-D and 2-D views)
represents the fitted model. The red segmentation (in the 2-D views)
is a contour plot of the data. The rendering and the segmentations
highlight the data and model at the $rms$ value chosen in the \textit{Contour level}
widget (Fig.~\ref{SlicerAstroFig5Widgets}).
In the chart view, the values of the ring parameters
of the fitted model are plotted (it is possible to switch the plots
in the chart view menu under the awl widget). 
The values of the parameters are also reported in 
the table widget on the $\tt{AstroModeling}$ module window interface
(see Fig.~\ref{SlicerAstroFig5Widgets}).
The values in the table are editable and can be used
to refine the model.
}
\label{SlicerAstroFig6}
\end{figure*} 

Figs.~\ref{SlicerAstroFig5} and \ref{SlicerAstroFig5Widgets} show
in blue a selection of the central body of WEIN069 and
the parameters chosen for running the fitting routine in $\tt{^{\rm3D}\,Barolo}$.
The fitting results are shown in Figs.~\ref{SlicerAstroFig5Widgets} and 
\ref{SlicerAstroFig6}: the yellow segmentation, in the 
2-D and 3-D views, represents the model, while the green rendering, in the
3-D view, represents the data. The visualization highlights the model and the data
at the  intensity level chosen in the \textit{contour level}
interface widget (in this case three times the value of the $rms$ noise in
the input data-cube).
 
The overlay of the segmentation of the model on the 3-D rendering of the data 
facilitates the inspection of the model. In the case of Fig.~\ref{SlicerAstroFig6},
the horizontal and vertical axes of the third 
3-D view (middle left panel) are the velocity and the 
declination dimensions, respectively. 
It is immediately clear that the
rotation curve of the model in the inner rings does not rise fast enough.
User interactions with the 3-D view such as camera zooming
and rotation enhance the 3-D perspective giving an even better overview of the differences.
On the other hand, for 
checking the data pixel by pixel (e.g., for data probing)
it is better to use a two-dimensional representation.

\begin{figure*}[!ht]
\centering
\includegraphics[width=1.\textwidth]{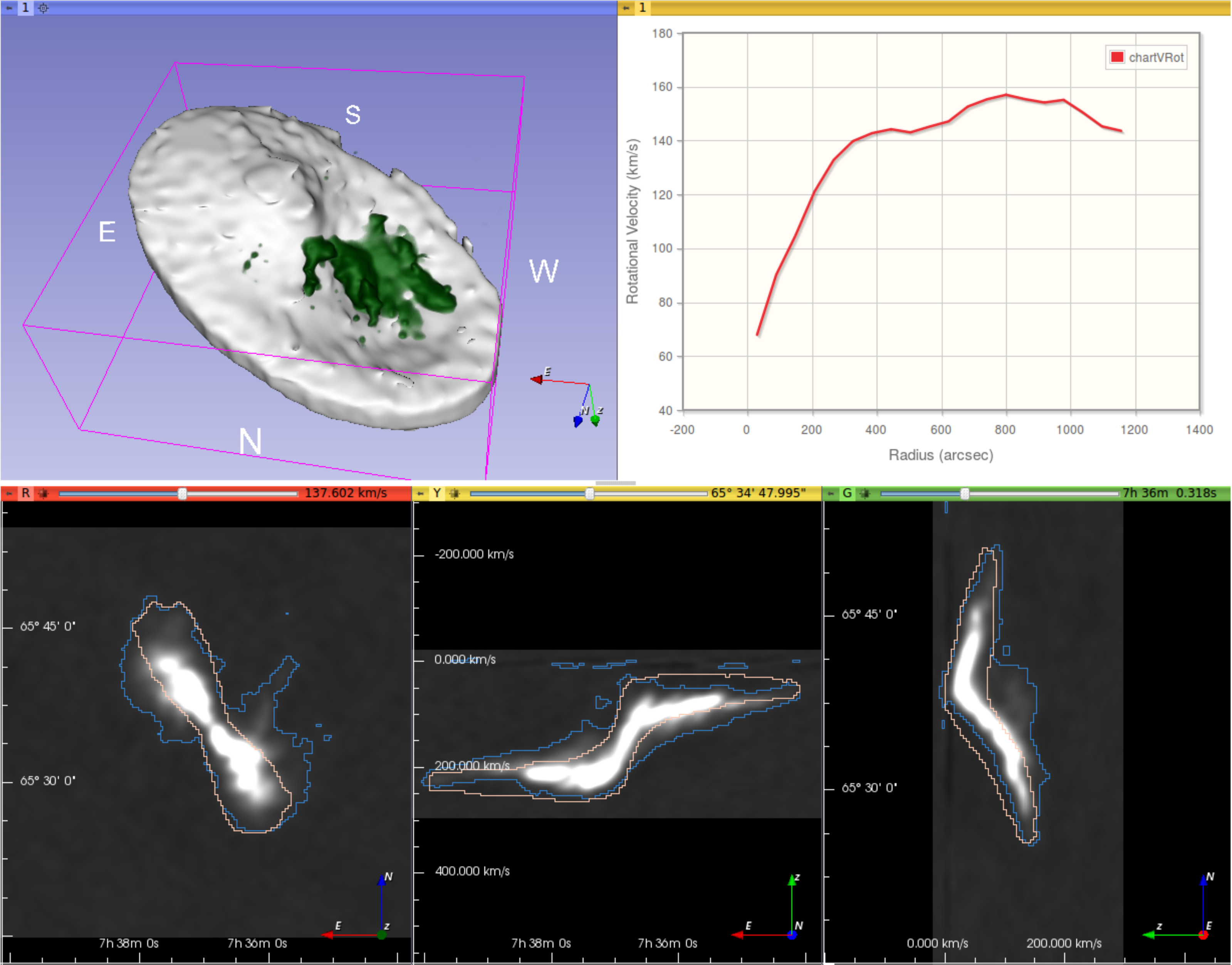}
\caption{Illustration of the  H\,{\scriptsize I} data and model, 
fitted by $\tt{^{\rm3D}\,Barolo}$,
of NGC2403 from the THINGS survey \citep{Walter}. The galaxy is very well resolved.
The comparative layout is the same as used in Fig.~\ref{SlicerAstroFig6}.
The top-left view is a 3-D view of NGC2403. The white labels 
represent the four cardinal directions (\textit{N}, \textit{S}, 
\textit{E}, \textit{W}). The green arrow points along the line of sight.
The white segmentation (in the 3-D view)
represents the model that fits the regular disk. 
The model has been fitted in \textit{automatic}
mode (i.e., no mask and no input parameters have 
been provided to $\tt{^{\rm3D}\,Barolo}$ 
from the graphical user interface).
The dark green rendering of the data, from an 
intensity level of 3 times the $rms$,
clearly shows unsettled gas in the inner region. 
The top-right view has plotting capabilities and the different parameters
of the rings of the output model can be shown. 
The bottom views are slices of the data-cube:
bottom-left, XY; bottom-middle, XZ; bottom-right, ZY.
The blue and the pink segmentations (in the 2-D views)
are contours of the data and model, respectively, at the $rms$ value chosen 
in the \textit{Contour level} widget (see Fig.~\ref{SlicerAstroFig5Widgets}).
}

\label{SlicerAstroFig7}
\end{figure*}

Finally, the $\tt{AstroModeling}$ module provides a table widget in the interface 
(see Fig.~\ref{SlicerAstroFig5Widgets}) that can be 
used to refine the model and update the visualization. 
All the ring parameters of the model
are available in the table. The refining process of the output model is 
crucial. In fact, the tilted-ring model 
fitting is a process with a high degree of degeneracy between the parameters,
and the fitting results strongly depend on the value of 
the initial parameters, especially for
the inclination. Therefore, the models must be carefully checked, compared
with the data, and refined.

The computational time needed by $\tt{^{\rm3D}\,Barolo}$ to fit the data depends on
several factors: the number of voxels, the number of rings of the model and the 
\textit{goodness} (i.e., whether the error in the estimate is $< 10\%$)
of the initial parameters. 
However, it is not possible to provide unbiased benchmarks for a fitting routine
(i.e., the performance highly depends on the input parameters).
To give an example, the time for fitting
a source extended up to $10^6$ voxels, using 20 rings and
having a reliable estimation of the input parameters is $\sim 2.5$ min 
(exploiting 1 CPU core at 2.60 GHz).
On the other hand, once the fitting has been performed, recalculating a single new model
with the same size in voxels and number of rings
takes less than 2 seconds (the process includes getting the model
from $\tt{^{\rm3D}\,Barolo}$ and creating the 
3-D segmentation in $\tt{SlicerAstro}$ as well). 
The computational complexity of this second step
is $O(Nr)$, where $N$ is the number of voxels of the source and $r$ the number of rings.
Despite the fact that the framework is not interactive, it is still fast enough to provide 
a powerful tool to refine models and compare them with the data.

\subsection{Use Case B: finding anomalous velocity gas}\label{UseCaseB}

It has been demonstrated that the gas distribution of some 
spiral galaxies \citep[e.g., NGC2403;][]{Fraternali}
is not composed of just a cold \textit{regular} thin disk. Stellar winds and 
supernovae can produce extra-planar gas \citep[e.g., a galactic fountain;][]{Bregman}. In this case, 
modeling is used to constrain the 3-D structure and kinematics of the extra-planar gas 
which is visible in the data as a faint kinematic component in addition to the disk.  

The $\tt{AstroModeling}$ module uses the output model of $\tt{^{\rm3D}\,Barolo}$
for visually highlighting the different components in the data-cube. 
After visualizing the model of the symmetric cold thin disk as a segmentation, it is immediately possible 
to locate any unusual features in the data-cube of interest and already get an idea of their 
properties, thus directing further modeling. For example, a model of the extra-planar gas above or 
below the disk with a slower rotation and a vertical motion provides quantitative information 
about the rotation and the infall velocity of such gas. 

In Fig.~\ref{SlicerAstroFig7} we show as an example the analysis that we performed on NGC2403.
The input parameters for the fitting have not been edited, therefore $\tt{^{\rm3D}\,Barolo}$
performed an automatic estimation of the initial parameters.
In the 3-D view $\tt{SlicerAstro}$ illustrates the data of the NGC2403 observations
rendered in green and the tilted-ring model generated with $\tt{^{\rm3D}\,Barolo}$
as a white segmentation. The white segmentation is rendered with the maximum opacity 
in order to obscure all the data that have been fitted by $\tt{^{\rm3D}\,Barolo}$.
This combination gives an immediate overview of the extra-planar gas present in 
the NGC2403 observations \citep{Fraternali}. Since $\tt{^{\rm3D}\,Barolo}$ 
mostly fits the symmetric regularly rotating part of the galaxy,
it therefore is a powerful tool for locating anomalous features
in the data, such as the extra-planar gas in NGC2403.

\section{Summary}\label{conclusion}
$\tt{SlicerAstro}$ is an open-source project and its binaries are currently
available in the extensions manager of
$\tt{3DSlicer}$\footnote{The user guide is available at the following link:
\url{https://github.com/Punzo/SlicerAstro/wiki\#get-slicerastro}}.  
The novelty of $\tt{SlicerAstro}$ over traditional 
astronomical viewers is the use of 3-D interactive tools
for the visualization and analysis of H\,{\small I} in and around galaxies.
$\tt{SlicerAstro}$ has been designed with
a strong, stable and modular \CC\; core, but it can be used also via $\tt{Python}$ scripting,
allowing great flexibility for user-customized visualization and analysis tasks 
(see Section~\ref{design}).

Although $\tt{SlicerAstro}$ is still under development, 
it already offers several new qualitative 
and quantitative visualization and analysis features 
for the inspection of H\,{\small I} and other spectral line data. 
The overall advantage of $\tt{SlicerAstro}$ compared to 
traditional viewers 
\citep[e.g., $\tt{KARMA}$, $\tt{Casaviewer}$ and $\tt{VISIONS}$;][]{Karma, CASA, Gipsy}
is that it bundles analytical operations such as smoothing
and modeling with the visualization.
These visual analytics techniques enhance the visualization itself. 
More important is in our view the interactivity offered by
$\tt{SlicerAstro}$. Interactivity is key to enhancing the
inspection and analysis of complex datasets. 
In fact, precisely the interactive and coupled 3-D/2-D visualization aspects 
(e.g., volume rendering, navigation, 
changing color/opacity function, selecting regions of interest
in the 3-D space) which are (partially) missing in the traditional tools 
and which disclose powerful visual analytics capabilities.

In Section~\ref{framework}, we presented the main module, $\tt{Astro}$-$\tt{Volume}$. 
This module provides: a user interface for loading and writing FITS files;
the display of astronomical World Coordinates; 
control of 2-D and 3-D color transfer functions; MRML nodes for
storing the data; and data conversion tools for masks and 
$\tt{3DSlicer}$ segmentation objects. Fig.~\ref{SlicerAstroFig1} showed 
how 3-D visualization gives an immediate overview of the H\,{\small I}
emission in and around WEIN069 
\citep[i.e.~three interacting galaxies and a tidal tail][]{wein,Mpati}
and the mask generated by automated source finder pipelines such as $\tt{SoFiA}$
or $\tt{Duchamp}$.
3-D visualization highly enhances and accelerates the inspection 
of the data and of the masks, allowing efficient manual quality control
of part (i.e., complex galaxies or groups of galaxies) of the large data 
sets that will be provided by the SKA precursors.

In addition, we presented the $\tt{AstroSmoothing}$ module in Section~\ref{filtering}.
Fig.~\ref{SlicerAstroFig2} showed the filtered version of WEIN069, 
obtained with a newly implemented intensity-driven gradient
filter \citep{Punzo2015}. 
The 3-D visualization highlights immediately the presence of a
faint filament between two galaxies that was hardly visible in the
original data-cube.
The coupling between the interactive smoothing algorithms 
(available in the parallelized version both on CPUs and GPUs) and
the 3-D visualization allows for a detailed inspection of the result
and a manual, iterative, search for the best smoothing 
parameters for maximally enhancing the local 
signal-to-noise ratio of the very faint signal.

Moreover, we introduced the $\tt{AstroCloudLasso}$ selection tool in Section~\ref{masking}.
This is a 3-D interactive selection tool \citep{Yu1}, optimized for H\,{\small I} data,
added by $\tt{Slicer}$-$\tt{Astro}$ in the $\tt{SegmentationEditor}$
of $\tt{3DSlicer}$. We showed how to use this tool to create and modify 
segmentation objects in the 3-D views (Fig.~\ref{SlicerAstroFig3}). 
The tool can be also used in the 2-D views for a
2-D selection. $\tt{CloudLasso}$ is an intuitive and efficient
3-D selection method, which is crucial for
allowing manual modification of
masks generated automatically by source finder pipelines 
(e.g., adding very faint signal missed by automated pipelines).
A second application of the tool is to select a region of interest (ROI). 
The ROI can be successively used to perform
calculations, such as tilted-ring model fitting, in the selection. 

In Section~\ref{modeling}, we demonstrated that 3-D visualization, 
coupled to modeling, provides additional capabilities 
helping the discovery and analysis of subtle structures in the 3-D domain.
We integrated $\tt{^{\rm3D}\,Barolo}$, a tilted-ring model fitting package,
in $\tt{SlicerAstro}$, providing 
an interface to set the input parameters 
for the fitting (Fig.~\ref{SlicerAstroFig5}). Moreover, 
the interface includes a widget for editing the ring parameters of the output
model, for recalculating and visualizing the model on top of the data. We also
showed that 3-D is a powerful tool not only to provide a region of interest
for the calculations, but also for 
the inspection of the model (Fig.~\ref{SlicerAstroFig6}) and 
the data not fitted by the model (e.g.~extra-planar gas in NGC2403, 
Fig.~\ref{SlicerAstroFig7}).

The efficiency and the effectiveness of the 
visual analytics techniques implemented in $\tt{SlicerAstro}$ have been tested. 
Quantifying the results for the efficiency of the modeling capabilities 
in $\tt{SlicerAstro}$ is not straightforward as the speed depends 
on the size of the data-cube and on the input parameters. However, even in the case
of a moderately large and well resolved object such as NGC2403 
(dimension $\sim 1.4 \times 10^6$ voxels), the
model fitting is performed in less than 2 minutes. 
In addition, modifying manually the parameters of the output model is interactive.
The 3-D smoothing 
algorithms in $\tt{SlicerAstro}$ have interactive performance. An adaptive
smoothing operation on NGC2403 data-cube is performed
in less than 0.1 seconds exploiting the computing power of a GPU (i.e., GeForce GTX860M).
The main advantage of $\tt{SlicerAstro}$ is that the combination of these 
smoothing and modeling capabilities with \textit{interactive} 3-D visualization provides an
immediate overview of all the coherent 3-D structures of the data, 
masks and models. This is very powerful and it definitely does increase 
the effectiveness of the visualization and, thus, the efficiency 
of the astronomical users in the manual exploration of many datasets.

We conclude that interactive 3-D quantitative and comparative 
visualization, 3-D user interaction and analysis capabilities available in 
$\tt{SlicerAstro}$ form an effective new tool that will boost, in terms both of
efficiency and quality, the analysis of complex sources in the context of
large data-flows that will be provided by the
SKA precursors. However, in order to fulfill all the visualization 
requirements defined in the Section~\ref{design}
\citep[and extensively discussed in][]{Punzo2015}, some quantitative
features still have to be incorporated in $\tt{SlicerAstro}$.
For example, a tool displaying the histogram of the flux intensities 
of the data-cube
will greatly help the user in setting the 2-D color function.
The capability to display flux density profiles (i.e.~linked 1-D visualization)
is also necessary, especially when dealing with unresolved sources.
In addition, capabilities for overlaying (in an automated way) other 
datasets (including datasets with different grids and 
projection systems) 
will enhance the inspection of multi-wave bands datasets and are 
under development. 
Furthermore, a dedicated tool in $\tt{SlicerAstro}$ for easily  
displaying position-velocity (P-V) diagrams will improve the
inspection and comparison of models. 
Specialized analysis tasks on 3-D selections (e.g., calculating statistics, 
moment maps, etc., in regions of interest) can be performed by 
running scripts in the
$\tt{3DSlicer}$ $\tt{Python}$ console. On the other hand, 
customized quantitative tasks can also be added, 
as core modules, in $\tt{SlicerAstro}$, similar to the 
implementation of the $\tt{AstroModeling}$ module.
These capabilities will be integrated in future updates of $\tt{SlicerAstro}$. 

The implementation of VO interoperability
and the advantages of such connectivity will be considered and analyzed
further in the case of $\tt{SlicerAstro}$. In fact, the SAMP protocol
and the FITS format are no longer globally accepted standards
\citep{Fits1,Fits2,Fits3}. Other scientific fields 
such as medical imaging can provide insights on how to improve the
astronomical standards.
For example, the Digital Imaging and Communications in Medicine (DICOM) \citep{dicom} protocol
is a remarkable example of a standard for handling, storing, printing,
and transmitting information in medical imaging. 
DICOM includes a file format definition and a network communications protocol
universally accepted and used by the medical scientific community.
 
$\tt{SlicerAstro}$ is a project under continuous development and
we have adopted an agile development approach (i.e.~development cycles 
are driven by user-feedback). In addition, the software is 
open-source and third parties are encouraged
to contribute. More important, any idea, feedback, criticism or bug
can be reported at the following link in the tracker
issue\footnote{\url{https://github.com/Punzo/SlicerAstro/issues}}. 

Finally, although the development of $\tt{SlicerAstro}$ thus far mainly
focused on 3-D H\,{\small I} data, it will also be a useful tool 
for any other type of 3-D astronomical data such as 
mm/submm molecular line data and 
optical integral field spectroscopic data. Molecular line data and 
optical/NIR spectroscopic data have the additional complication that
often more than one spectral line are present in a single spectral 
window. This makes the visualization more complex, though clever 
stacking of the known spectral lines can e.g.~enhance the signal to
noise and stacking tools can help to bring out the kinematic behavior 
of gas which emits multiple spectral lines.  One can also think of 
additional tools to visualize line ratios by e.g.~superposing different spectral
lines in different colors interactively. In conclusion, though
$\tt{SlicerAstro}$ is useful for other types of astronomical 3-D data, additional tools 
will be required tailored to the kind of data and the scientific 
questions to be addressed.

\section{Appendix}\label{Appendix}

Below, the $\tt{Python}$ code provides an example for 
applying smoothing to a data-cube, performing the rendering and 
saving the last as a video. The script can be copy and pasted in   
the $\tt{3DSlicer}$ $\tt{Python}$ console, or it can be launched
from the command line with the following command:

\begin{lstlisting}[language=bash]
./Slicer --python-script script.py
\end{lstlisting}

\textcolor{blue}{More information is provided at the following link:}
\url{https://github.com/Punzo/SlicerAstro/wiki}.

\begin{python}

# Load a data-cube in SlicerAstro
slicer.util.loadVolume("/full_path/WEIN069.fits",{"center":True})
mw = slicer.util.mainWindow()
ms = mw.moduleSelector()

# Smooth the data-cube in automatic mode (CPU)
ms.selectModule('AstroSmoothing')
smowidget = slicer.modules.astrosmoothing.widgetRepresentation()
smowidget.onApply()

# Setup the Render for the data-cube and its filtered version
ms.selectModule('AstroVolume')
astrovolumewidget = slicer.modules.astrovolume.widgetRepresentation()
astrovolumewidget.onCurrentQualityControlChanged(1)
volumes = slicer.mrmlScene.GetNodesByClass("vtkMRMLAstroVolumeNode")
volumefiltered = volumes.GetItemAsObject(1)
smomrmlpara.SetInputVolumeNodeID(volumefiltered.GetID())
astrovolumewidget.onCurrentQualityControlChanged(1)

# Create videos 
ms.selectModule('ScreenCapture')
screencapturewidget = slicer.modules.screencapture.widgetRepresentation()
instance = screencapturewidget.self()
# For the data-cube
viewNode = slicer.util.getNode('vtkMRMLViewNode1')
instance.viewNodeSelector.setCurrentNode(viewNode)
instance.numberOfStepsSliderWidget.setValue(360)
instance.videoExportCheckBox.setChecked(1)
instance.videoFormatWidget.setCurrentIndex(1)
instance.videoFileNameWidget.setText("WEIN069.mp4")
instance.videoLengthSliderWidget.setValue(6)
instance.onCaptureButton()
# For the filtered version
viewNode = slicer.util.getNode('vtkMRMLViewNode2')
instance.viewNodeSelector.setCurrentNode(viewNode)
instance.numberOfStepsSliderWidget.setValue(360)
instance.videoExportCheckBox.setChecked(1)
instance.videoFormatWidget.setCurrentIndex(1)
instance.videoFileNameWidget.setText("WEIN069_smoothed.mp4")
instance.videoLengthSliderWidget.setValue(6)
instance.onCaptureButton()

\end{python}

\section{Acknowledgments}
We thank M.A. Ramatsoku and M.A.W. Verheijen 
for proving us with the H\,{\small I} data of WEIN069. 
Support also came from S. Pieper (Isomics, Inc.),
A. Lasso (Laboratory of Percutaneous Surgery at Queen's University) 
and E. di Teodoro (Australian National University)
in the form of feedback and assistance.

Finally, we thank the reviewers for their constructive
comments, which helped us to improve the paper.

D. Punzo and J.M van der Hulst acknowledge the support from the European 
Research Council under the European Union's Seventh
Framework Programme (FP/2007-2013)/ERC 
Grant Agreement nr. 291-531. 

L. Yu was partially supported from National Natural 
Science Foundation of China (Grant No. 61502132).

We are grateful to the various agencies and programs that funded
support and development of $\tt{3DSlicer}$ over the years.

\section{References}
\bibliographystyle{elsart-num-names}
\bibliography{SlicerAstro.bib}

\end{document}